\documentclass[aps,prd,twocolumn,showpacs,10pt,superscriptaddress,preprintnumbers,nofootinbib]{revtex4-1}
\usepackage{epsfig,amssymb,amsmath,psfrag,epstopdf,color}
\pdfoutput=1
\usepackage{graphicx}
\usepackage{subfig}
\usepackage[margin=7pt,justification=centerlast]{caption}
\usepackage[bookmarks=true]{hyperref} 
 
\usepackage{paralist}
\usepackage{hyperref}
\allowdisplaybreaks

\def\beq{\begin{equation}}
\def\eeq{\end{equation}}
\def\bsp#1\esp{\begin{split}#1\end{split}}
\newcommand{\be}{\begin{equation}}
\newcommand{\ee}{\end{equation}}
\newcommand{\bea}{\begin{eqnarray}}
\newcommand{\eea}{\end{eqnarray}}

\def\Fig#1{Fig.~{\ref{#1}}}

\def\cN{{\mathcal N}}
\def\cO{{\mathcal O}}
\def\cL{{\mathcal L}}

\def\to{\rightarrow}

\def\ksl{\not{\hbox{\kern-2.3pt $k$}}}

\def\Ord{{\cal O}}

\def\eqn#1{Eq.~(\ref{#1})}

\def\spa#1.#2{\left\langle#1\,#2\right\rangle}
\def\spb#1.#2{\left[#1\,#2\right]}
\def\lor#1.#2{\left(#1\,#2\right)}
\def\sand#1.#2.#3{%
\left\langle\smash{#1}{\vphantom1}^{-}\right|{#2}%
\left|\smash{#3}{\vphantom1}^{-}\right\rangle}

\newcommand{\fd}[2]{\parbox{#1}{\includegraphics[width=#1]{#2}}}

\newcommand{\nn}{\nonumber}

\begin{document}

\preprint{SLAC--PUB--17427}

\title{The Collinear Limit of the Energy-Energy Correlator}
\author{Lance~J.~Dixon}
\affiliation{SLAC National Accelerator Laboratory, Stanford University, CA, USA\vspace{0.5ex}}
\author{Ian~Moult}
\affiliation{Berkeley Center for Theoretical Physics, University of California, Berkeley, CA 94720, USA\vspace{0.5ex}}
\affiliation{Theoretical Physics Group, Lawrence Berkeley National Laboratory, Berkeley, CA 94720, USA\vspace{0.5ex}}
\author{Hua~Xing~Zhu}
\affiliation{Zhejiang Institute of Modern Physics, Department of
  Physics, Zhejiang University, Hangzhou, 310027, China\vspace{0.5ex}}

\begin{abstract}
The energy-energy-correlator (EEC) observable in $e^+e^-$ annihilation measures the energy deposited in two detectors as a function of the angle between the detectors.  The collinear limit, where the angle between the two detectors approaches zero, is of particular interest for describing the substructure of jets produced at hadron colliders as well as in $e^+e^-$ annihilation.  We derive a factorization formula for the leading power asymptotic behavior in the collinear limit of a generic quantum field theory, which allows for the resummation of logarithmically enhanced terms to all orders by renormalization group evolution. The relevant anomalous dimensions are expressed in terms of the timelike data of the theory, in particular the moments of the timelike splitting functions, which are known to high perturbative orders. We relate the small angle and back-to-back limits to each other via the total cross section and an integral over intermediate angles.  This relation, for the EEC in $e^+e^-$ and in Higgs decay to gluons, provides us with the initial conditions for quark and gluon jet functions at order $\alpha_s^2$.  In QCD and in $\cN=1$ super-Yang-Mills theory, we then perform the resummation to next-to-next-to-leading logarithm, improving previous calculations by two perturbative orders.  We highlight the important role played by the non-vanishing $\beta$ function in these theories, which while subdominant for Higgs decays to gluons, dominates the behavior of the EEC in the collinear limit for $e^+e^-$ annihilation, and in $\cN=1$ super-Yang-Mills theory. In conformally invariant $\cN=4$ super-Yang-Mills theory, reciprocity between timelike and spacelike evolution can be used to express our factorization formula as a power law with exponent equal to the spacelike twist-two spin-three anomalous dimensions, thus providing a connection between timelike and spacelike approaches.
\end{abstract}

\maketitle

\section{Introduction}
\label{sec:introduction}

\begin{figure}[h]
  \centering \includegraphics[origin=c,width=0.3\textwidth]{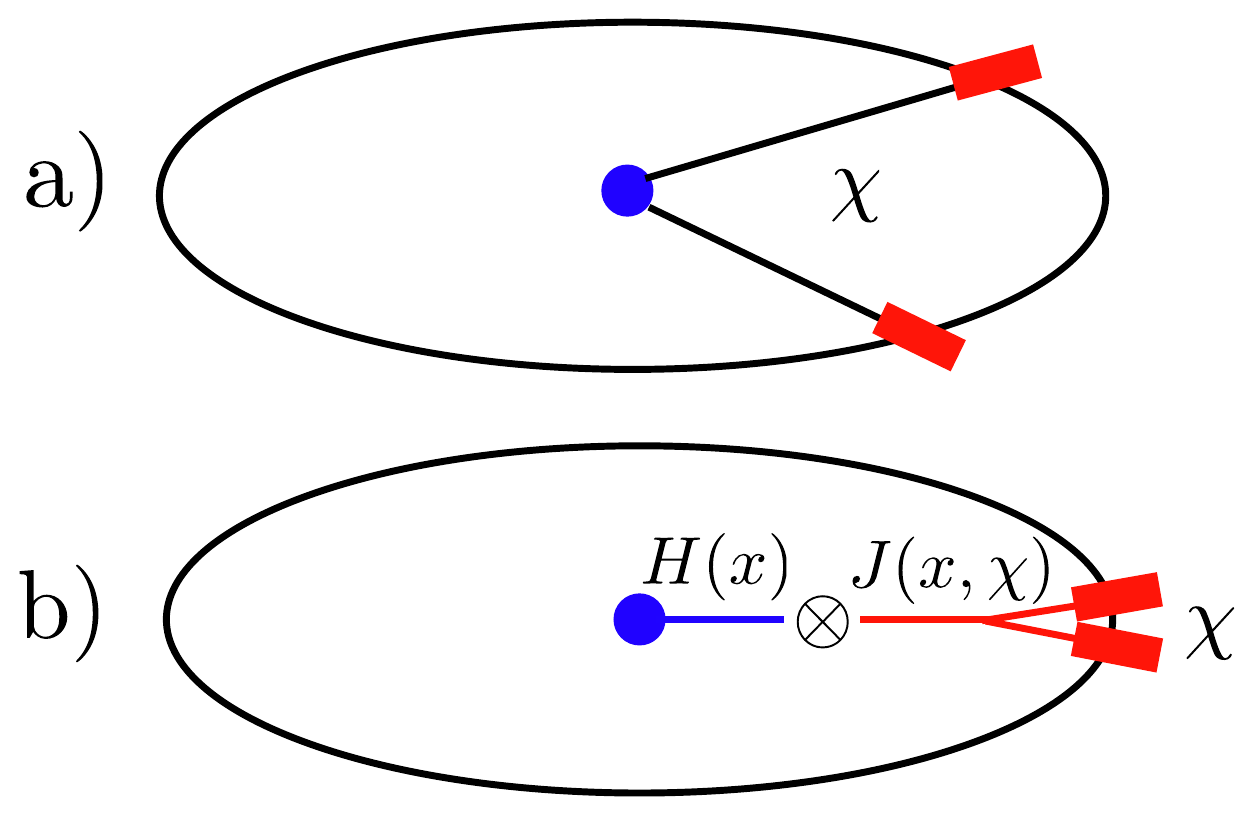}
  \caption{a) \!The EEC observable for a generic angle \!$\chi$. b) In the collinear limit the EEC factorizes into a hard function, $H(x)$, describing the production of a parton of momentum fraction $x$ from the source, and a collinear jet function, $J(x,\chi)$, describing the measurement.}
  \label{fig:def}
\end{figure}

Jet and event shape observables play a crucial role in our understanding of QCD, and are interesting more generally for understanding the structure of Lorentzian observables in quantum field theory.  A particularly interesting infrared-safe observable is the energy-energy correlator (EEC), originally defined in $e^+e^-$ annihilation~\cite{Basham:1978bw,Basham:1978zq}, which measures the energy in two detectors separated by an angle $\chi$, see \Fig{fig:def}.  The EEC can be defined within QCD also for a gluonic source, namely the decays of a Higgs boson to hadrons that are mediated by a heavy top quark loop~\cite{Luo:2019nig}.  The EEC has also been studied in conformally invariant $\cN=4$ super-Yang-Mills theory (SYM) for sources that are protected by supersymmetry~\cite{Hofman:2008ar,Belitsky:2013xxa,Belitsky:2013bja,Belitsky:2013ofa}. It exhibits kinematic singularities in both the back-to-back ($\chi\to \pi$) and collinear ($\chi\to 0$) limits, allowing its behavior in these limits to be understood to all orders in perturbation theory using renormalization group techniques. The compatibility of these two limits suggests a particularly rigid structure, perhaps enabling an all orders perturbative understanding of the EEC. 

The EEC has attracted significant recent attention, which has further revealed its perturbative simplicity. Advances include analytic results for arbitrary $\chi$ to next-to-leading order (NLO) in QCD \cite{Dixon:2018qgp,Luo:2019nig} and at both NLO \cite{Belitsky:2013ofa} and NNLO \cite{Henn:2019gkr} in $\cN=4$ SYM; an understanding of the all orders logarithmic structure in the back-to-back limit $\chi\to\pi$ \cite{Moult:2018jzp,Gao:2019ojf}; and numerical results at NNLO in QCD \cite{DelDuca:2016ily} that have been matched~\cite{Tulipant:2017ybb} to the next-to-next-to-leading logarithms (NNLL) in the back-to-back limit~\cite{deFlorian:2004mp} and used to determine the strong coupling~\cite{Kardos:2018kqj}.

Recently a description of the all-orders behavior in the collinear limit for a conformal field theory has been given~\cite{DavidPetrSashaTalks,DSDtocome} based on the light-ray operator formalism~\cite{Kravchuk:2018htv,Kologlu:2019bco}. The limit is described by a spacelike operator product expansion (OPE) controlled by the twist-two spin-three operator whose role was identified earlier~\cite{Konishi:1979cb,Hofman:2008ar}.  Another spacelike approach to the collinear limit in a CFT has been developed more recently~\cite{Korchemskytocome}, based on the representation of the EEC in terms of the Mellin amplitude of the four-point function~\cite{Belitsky:2013xxa,Belitsky:2013bja,Belitsky:2013ofa}.

Despite this progress, the all orders logarithmic structure in the collinear limit remains less well understood for a generic quantum field theory. The leading logarithms (LL) in the $\chi\to 0$ limit have been resummed to all orders in QCD using the jet calculus approach \cite{Konishi:1978yx,Konishi:1978ax,Konishi:1979cb,Kalinowski:1980wea,Richards:1982te}. However, there has not been a systematic framework for resumming subleading logarithms. In addition to being of formal interest, the collinear limit is particularly relevant for the study of jets and their substructure at the Large Hadron Collider (LHC), motivating an improved quantitative understanding.

In this paper we present a factorization formula describing the  $\chi\to 0$ limit of the EEC in a generic massless quantum field theory, conformal or asymptotically free. All logarithms in the perturbative expansion can be resummed using the renormalization group evolution of certain jet functions appearing in the factorization formula. We show that the anomalous dimensions of these functions are related to the timelike twist-two anomalous dimensions governing the evolution of fragmentation functions for identified hadrons.  These timelike splitting kernels, along with the corresponding hard functions or matching coefficients, are known through NNLO in QCD~\cite{Rijken:1996ns,Mitov:2006wy,Mitov:2006ic,Moch:2007tx,Almasy:2011eq}.  These results facilitate the determination of the asymptotic behavior of the EEC in the $\chi\to 0$ limit to high perturbative orders.  We explicitly resum the EEC to NNLL accuracy in QCD and in $\cN=1$ SYM, improving by two logarithmic orders the best known results in the literature. In the particular case of $\cN=4$ SYM, a reciprocity that relates timelike and spacelike anomalous dimensions \cite{Drell:1969jm,Gribov:1972ri,Mueller:1983js,Blumlein:2000wh,Dokshitzer:2005bf,Marchesini:2006ax,Basso:2006nk,Dokshitzer:2006nm} allows us to express our result as a power law, where the exponent is the twist-two spin-three spacelike anomalous dimension \cite{Hofman:2008ar}.  This relation provides a link between timelike dynamics and spacelike data.

An outline of this paper is as follows. In Sec.~\ref{sec:obs_def} we review the definition of the EEC observable. In Sec.~\ref{sec:formalism} we present our factorization formula for the collinear limit of the EEC. In Sec.~\ref{sec:both_ends} we discuss a sum rule arising from the overall normalization of the cross section and how this enables us to obtain the two loop jet function for the EEC. In Secs.~\ref{sec:NNLL},~\ref{sec:landau} and~\ref{sec:N4} we study the behavior of the collinear limit of the EEC in QCD, $\cN=1$ SYM and $\cN=4$ SYM, highlighting several interesting features of each case. We conclude in Sec.~\ref{sec:conclusion}, and discuss a number of interesting future directions.  We also provide an ancillary file supplying an iterative solution through nine loops to the NNLL jet function evolution equations in QCD.

\section{Observable Definition}
\label{sec:obs_def}

The EEC is defined as \cite{Basham:1978bw}
\begin{align}
  \label{eq:EECdef}
  \frac{d\sigma}{dz}= \sum_{i,j}\int d\sigma\ \frac{E_i E_j}{Q^2} \delta\left(z - \frac{1 - \cos\chi_{ij}}{2}\right) \,,
\end{align}
where $d\sigma$ is the product of the squared matrix element and the phase-space
measure, $E_i$ and $E_j$ are the energies of final-state partons $i$ and $j$
in the center-of-mass frame, and their angular separation is $\chi_{ij}$.
For convenience, we have chosen to work with the variable $z$ satisfying
\begin{align}
0\leq z=\frac{1-\cos \chi}{2} \leq 1\,.
\end{align}
Due to the fact that $Q^2 = (\sum_i E_i)^2 = \sum_{i,j} E_i E_j$, the EEC observable satisfies the normalization condition
\begin{align}
  \label{eq:sumrule}
  \int_0^1 dz\, \frac{d\sigma}{dz}= \sigma_{\rm tot} \,.
\end{align}
As we will see in Sec.~\ref{sec:both_ends}, this relation places strong constraints on the cross section, and in particular, links the singular behavior at the two kinematic endpoints.

In the collinear limit, $z \to 0$, the perturbative contributions to the EEC exhibits a single logarithmic series
\begin{align}
  \frac{d\sigma}{dz}= \sum_{L=1}^\infty \sum_{j=-1}^{L-1} \left(\frac{\alpha_s(\mu)}{4 \pi} \right)^L c_{L,j} \cL^j (z) +\ldots \,,  
\end{align}
where $\cL^{-1}(z)=\delta(z)$ and $\cL^{j}(z)= \left[\ln^j z/z \right]_+$ for $j\geq 0$ denotes a standard plus distribution.  The ellipses denote terms with a less singular power than $1/z$. (Note that $\delta(z) \sim 1/z$.) One of our primary goals will be to describe this logarithmic structure to all orders.

\section{Factorization Formula}
\label{sec:formalism}

It is convenient to work in terms of the cumulant of the EEC,
\begin{align}
  \label{eq:cumulant}
\Sigma\Bigl(z, \ln \frac{Q^2}{\mu^2} , \mu\Bigr)\
\equiv\ \frac{1}{\sigma_0}
\int^z_0 dz' \, 
\frac{d\sigma}{dz} \Bigl(z', \ln\frac{Q^2}{\mu^2}, \mu\Bigr) \,,
\end{align}
where $\sigma_0$ is the Born-level total cross section.
The cumulant maps $\left[\ln^j z/z \right]_+\ \to\ 1/(j+1)\times\ln^{j+1} z$ and $\delta(z)\ \to 1$.
The $\mu$-dependence in the last arguments of $\Sigma$ and $d\sigma/dz$ is entirely through the strong coupling $\alpha_s(\mu)$; we just write it as $\mu$ to save space.  One of the main results of this paper is a factorization formula for $\Sigma$ in the $z\to 0$ limit
\begin{align}
\label{eq:fact}
 \Sigma(z, \ln \frac{Q^2}{\mu^2} , \mu)
= \int_0^1 dx\, x^2 \vec{J} (\ln\frac{z x^2 Q^2}{\mu^2},\mu)
   \cdot  \vec{H} (x,\frac{Q^2}{\mu^2},\mu) \,.
\end{align}
This formula factorizes the dynamics in the collinear limit into a hard function $H$, which describes the dynamics of the source, but is independent of the measurement, $z$, and a jet function, $J$, which describes the $z$ dependence, but is independent of the source. This is illustrated in \Fig{fig:def}. Both the hard function and jet function are vectors in flavor space. For the particular case of QCD, where we have quarks and gluons, we have $\vec{J} = \{ J_q, J_g\}$ and $\vec{H} = \{H_q, H_g\}$. It is not necessary to distinguish $q$ and $\bar{q}$ due to the charge conjugation invariance of QCD and the symmetry of the source. Corrections to this factorization formula are suppressed by an integer power of $z$, as can be shown from the known structure of higher twist distribution functions \cite{Jaffe:1991ra}. %

The jet functions are gauge invariant non-local operators.
The quark jet function is defined as
\begin{align}
J_q(z)=\sum\limits_{X} \sum\limits_{i,j\in X}  \langle 0 | \bar \chi_n |X \rangle  \frac{E_i E_j}{(Q/2)^2} \Theta(\theta_{ij}<\chi) \langle X|\chi_n |0\rangle\,,
\end{align}
where $\chi_n$ is a gauge invariant collinear quark field in SCET \cite{Bauer:2000ew, Bauer:2000yr, Bauer:2001ct, Bauer:2001yt}. The $\Theta$ function on the parton separation angle $\theta_{ij}$ is appropriate for the cumulant definition of $\vec{J}$ in \eqn{eq:fact}.  The gluon jet function is defined in a similar manner, using a gauge invariant gluon field. (In a more general context, $Q/2$ would be replaced by the jet energy in an appropriate frame.)

The jet and hard functions both satisfy renormalization group (RG) evolution equations which allow for the resummation of logarithms of $z$. The RG equation for the hard function is given by
\begin{align}
  \label{eq:hardRG}
  \frac{d \vec{H} (x, \frac{Q^2}{\mu^2},\mu)}{d \ln \mu^2} = - \int_x^1 \frac{dy}{y} \widehat P_T(y,\mu) \cdot \vec{H}\left(\frac{x}{y}, \frac{Q^2}{\mu^2},\mu\right) \,,
\end{align}
where $\widehat P_T$ is the singlet timelike splitting kernel matrix
\begin{align}
  \label{eq:splitK}
  \widehat P_T = 
  \begin{pmatrix}
    P_{qq} &  P_{qg}
\\
    P_{gq} & P_{gg}
  \end{pmatrix} \,.
\end{align}
The jet function obeys the RG equation
\begin{align}
  \label{eq:jetRG}
\frac{d \vec{J}(\ln\frac{z Q^2}{\mu^2}, \mu) }{d \ln \mu^2} = \int_0^1 dy\, y^2 \vec{J} (\ln\frac{z y^2 Q^2}{\mu^2}, \mu) \cdot \widehat P_T(y,\mu) \,.
\end{align}
This equation can be derived by requiring the cumulant $\Sigma$ in \eqn{eq:fact} to be RG invariant, combined with the evolution equation~\eqref{eq:hardRG} for the hard function.

As indicated in \eqn{eq:fact}, logarithms in the jet function are minimized at the scale $\mu^2=z x^2 Q^2\equiv q_T^2$, which physically corresponds to a transverse momentum scale $q_T \approx \chi x Q/2$ associated with the splitting at momentum $xQ$ and angle $\chi$ measured by the EEC. The logarithms of the hard function are minimized at the scale $\mu^2=Q^2$, which corresponds to the energy scale of the source.  Resummation is achieved by computing the boundary values of the jet and hard functions at these scales, and then performing the RG evolution from one scale to the other.

The factorization formula in \eqn{eq:fact} is more complicated than the standard jet calculus formula which describes the leading logarithms~\cite{Konishi:1978yx,Konishi:1978ax,Konishi:1979cb,Kalinowski:1980wea,Richards:1982te}, due to the presence of the convolution in the momentum variable $x$. This convolution is only required beyond LL; at LL it suffices to set $x=1$ in the argument of $\vec{J}$. The evolution equation~\eqref{eq:jetRG} then simplifies to a multiplicative renormalization,
\begin{align}
  \label{eq:jetLL}
  \frac{d \vec{J}_{\rm LL}(\ln\frac{z Q^2}{\mu^2}, \mu) }{d \ln \mu^2}  =&
\, \vec{J}_{\rm LL}(\ln\frac{z Q^2}{\mu^2}, \mu)
\cdot \int_0^1 dy\, y^2 \, \widehat{P}_T^{(0)}(y)
\nn\\
=&\, - \vec{J}_{\rm LL}(\ln\frac{z Q^2}{\mu^2}, \mu)  \cdot \gamma_T^{(0)} \,,
\end{align}
where $\gamma_T \equiv \gamma_T(3)$ is the $N=3$ moment of the LO timelike singlet splitting kernel. At LO, the timelike and spacelike moments are identical, and are given by
\begin{align}
  \label{eq:gT0}
  \gamma_T^{(0)} = 
  \begin{pmatrix}
    \tfrac{25}{6} C_F &  - \tfrac{7}{15} n_f
\\
    - \tfrac{7}{6} C_F & \tfrac{14}{5} C_A + \tfrac{2}{3} n_f
  \end{pmatrix} \,.
\end{align}
We adopt the conventions of
refs.~\cite{Mitov:2006wy,Mitov:2006ic,Moch:2007tx,Almasy:2011eq}
for splitting kernels and anomalous dimensions, which are related by
a Mellin transform,
\begin{equation}
\label{eq:gammaPTrelation}
\gamma_T(N)\ \equiv\ - \int_0^1 dy \, y^{N-1} \, \widehat{P}_T(y).
\end{equation}
We also use the perturbative expansion parameter $a_s \equiv \alpha_s/(4\pi)$.

An exact solution to \eqn{eq:jetLL} is given by
\begin{align}
  \label{eq:jetLLresum}
  \vec{J}_{\rm LL}(\ln\frac{z Q^2}{\mu^2}, \mu) = (1,1) \cdot V \left[\left(\frac{\alpha_s(\sqrt{z} Q)}{\alpha_s(\mu)} \right)^{-\frac{\vec{\gamma}_T^{(0)}}{\beta_0}}\right]_D \! V^{-1} \,,
\end{align}
where $\beta_0 = (11C_A-2n_f)/3$, $V$ is the matrix that diagonalizes $\gamma_T^{(0)}$, and $\vec{\gamma}_T^{(0)}$ is the diagonal vector of the diagonalized matrix. Substituting this solution into \eqn{eq:fact}, using that
\begin{align}
  \label{eq:HLL}
  \vec{H}_{\rm LL} (x) = 
  \begin{pmatrix}
    \tfrac{1}{2} \delta(1-x)
\\
    0
  \end{pmatrix} \,,
\end{align}
and differentiating $\Sigma$ to obtain $d\sigma/dz$, we reproduce the LL resummation formula obtained using jet calculus. Beyond LL, the convolution in the momentum fraction variable, $x$, cannot be eliminated. Indeed, we will see in Sec.~\ref{sec:N4} that this convolution is crucial to obtain a correspondence with the spacelike picture in a conformal field theory (CFT).

\section{Jet Functions and Sum Rules}
\label{sec:both_ends}

The hard function and the timelike splitting kernel entering our factorization formula are known in QCD to NNLO \cite{Mitov:2006wy,Mitov:2006ic,Moch:2007tx,Almasy:2011eq}, however, the EEC jet functions are new. They can be computed from their operator definition, which at NLO is equivalent to integrating the splitting functions against the EEC measurement function. One subtlety when computing the jet functions is that the EEC detectors can both be placed on the same particle. This is in fact essential to obtain an IR finite jet function. Representative one-loop diagrams for the quark jet functions are
\begin{align}\label{eq:jet_diagram}
\ \raisebox{0.0cm}{\fd{1.85cm}{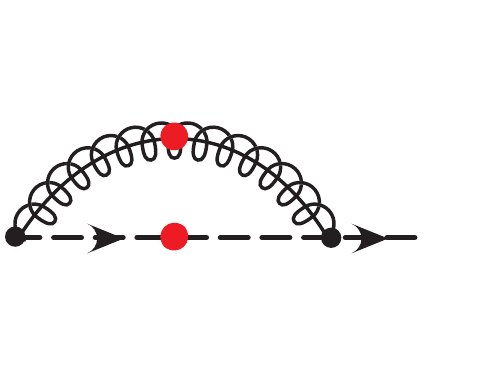}}\ +\ \raisebox{0.0cm}{\fd{2.15cm}{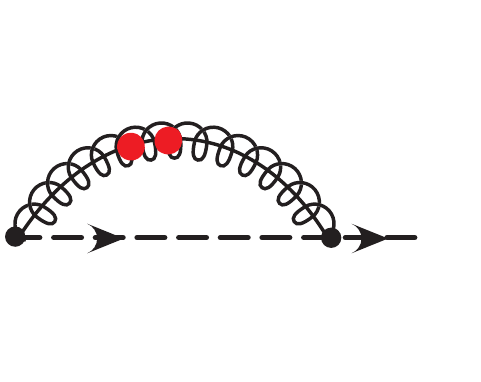}}+\ \raisebox{0.0cm}{\fd{2.15cm}{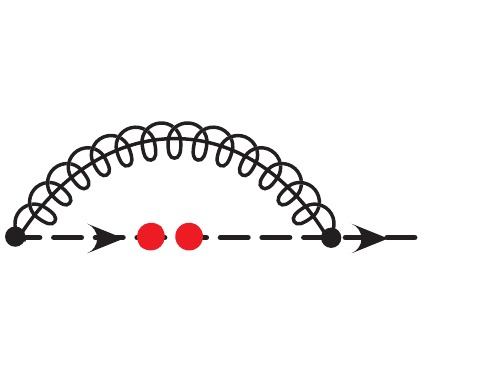}}\,,
\end{align}
where the red dots denote insertions of the EEC operators. Writing the perturbative expansion of the jet functions as $J_{q,g}=\sum_L a_s^L J^{(L)}_{q,g}$, with $J_{q,g}^{(0)} = 1$, a simple calculation gives the one-loop quark jet function,
\begin{align}
  \label{eq:oneloopquarkjet}
  J_q^{(1)} 
=& \, 3 C_F \left( - \frac{1}{\epsilon_{\rm UV}} + \ln \frac{z Q^2}{\mu^2}\right) + j_1^q  + \Ord(\epsilon), \\
j_1^q =& \, - \frac{37}{3} C_F \,.
\label{eq:oneloopquarkjetconstant}
\end{align}
Renormalization leads to mixing between the quark and gluon jet functions.
The one-loop gluon jet function can be computed in a similar manner; the result is
\begin{align}
  \label{eq:oneloopgluonjet}
  J_g^{(1)} =& \, \left( \frac{14}{5} C_A + \frac{1}{5} n_f \right) 
\left( - \frac{1}{\epsilon_{\rm UV}} + \ln \frac{z Q^2}{\mu^2}\right)
 + j_1^g  + \Ord(\epsilon), \\
j_1^g =& \, - \frac{898}{75} C_A - \frac{14}{25} n_f \,.
\label{eq:oneloopgluonjetconstant}
\end{align}
The pole and $\ln(zQ^2/\mu^2)$ coefficient are again dictated by the anomalous dimensions, here $\gamma_{qg}^{(0)}+\gamma_{gg}^{(0)}$.

The direct perturbative calculation of the jet function at NNLO is non-trivial due to the appearance of triple collinear splitting functions~\cite{Campbell:1997hg,Catani:1999ss} and the constraints on the three-particle phase space. Instead of performing a direct calculation, we can obtain the jet function by exploiting the sum rule~\eqref{eq:sumrule}.  Using the sum rule at ${\cal O}(\alpha_s^2)$ requires knowledge of the singular behavior in the back-to-back limit \cite{deFlorian:2004mp,Moult:2018jzp} and the analytic form of the NLO EEC for both $e^+e^-$ annihilation~\cite{Dixon:2018qgp} and hadronic Higgs decay~\cite{Luo:2019nig}.  It also needs the perturbative corrections to the total cross section, which are known in QCD to ${\cal O}(\alpha_s^4)$~\cite{Baikov:2012er,Herzog:2017dtz}.

To illustrate this idea, we recompute the NLO jet constants using this sum rule. The LO EEC in $e^+e^-$, including its end-point contributions, is given by
\begin{widetext}
\begin{align}
\frac{1}{\sigma_0} \frac{d \sigma(z,\mu=Q)}{dz}\Biggr|_{a_s^1}
=& \, \Bigl[ \frac{1}{2} j_1^q + h_1^q + h_1^g \Bigr] \delta(z)
+ C_F \biggl\{ ( - 2 \zeta_2 - 4 ) \delta(1-z)
+ \frac{3}{2} \frac{1}{[z]}_+
- 2 \left[\frac{\ln(1-z)}{1-z} \right]_+
- \frac{3}{[1-z]}_+
\nn\\
& \, + \frac{1}{2 z^5} \left[ -9 z^4-6 z^3-42 z^2 +36 z
   +4 ( -z^4-z^3+3 z^2-15 z+9 )
   \ln (1-z)\right] \biggl\} \,.
\end{align}
\end{widetext}
The factorization formula~\eqref{eq:fact} provides the $\delta(z)$ term, where $h_1^q = 131/16\,C_F$ and $h_1^g = - 71/48\,C_F$ are the $N=3$ moments of the NLO quark and gluon hard functions (normalized to be half the sum of the $T$ and $L$ angular coefficient functions in ref.~\cite{Mitov:2006wy}, as explained in App.~\ref{app}). The one-loop result for the total cross section is $\int_0^1 dz \, d\sigma/dz =  3 C_F \sigma_0 a_s$.  The bulk integral, defined to be the integral omitting the delta functions and plus distributions (the latter integrate to zero), is
\begin{equation}
\label{eq:QCDbulkLO}
\frac{1}{\sigma_0} \int_0^1 dz \frac{d\sigma^{{\rm bulk}}}{dz}\Bigg|_{a_s^1}
= C_F \left( 2 \zeta_2 + \frac{155}{24} \right) \,.
\end{equation}
Combining these results, we can extract $j_1^q = -37/3\,C_F$, which agrees precisely with \eqn{eq:oneloopquarkjetconstant}. Note that this computation requires the knowledge of the $\delta(1-z)$ term.

In order to extract the two two-loop jet function constants, we integrated the NLO EEC bulk cross sections for $e^+e^-$ and Higgs~\cite{Dixon:2018qgp,Luo:2019nig} numerically to high accuracy and reconstructed the result in terms of $\zeta$ values using the PSLQ algorithm. The result is
\begin{widetext}
\begin{align}
\frac{1}{\sigma_0} \int_0^1 dz \frac{d\sigma^{{\rm bulk}}_{e^+e^-}}{dz}\Bigg|_{a_s^2}
=&\ C_F n_f \left( \frac{8}{3} \zeta_3 - \frac{457}{180} \zeta_2
               - \frac{3016223}{216000} \right)
+ C_F C_A \left( 30 \zeta_4 - \frac{422}{3} \zeta_3 + \frac{893}{45} \zeta_2
               + \frac{19871011}{162000} \right) \nn\\
&
+ C_F^2 \left( - 92 \zeta_4 + 164 \zeta_3 - \frac{697}{12} \zeta_2
               - \frac{286843}{5184} \right) \,, 
\label{eq:QCDbulkNLO} \\
\frac{1}{\sigma_0} \int_0^1 dz \frac{d\sigma^{{\rm bulk}}_{H}}{dz}\Bigg|_{a_s^2}
=&\ n_f^2 \left( - \frac{6}{5} \zeta_2 + \frac{4371}{500} \right)
+ C_F n_f  \left( - \frac{104}{15} \zeta_3 + \frac{23}{10} \zeta_2
                 - \frac{42509}{12000} \right)
\nn\\
& + C_A n_f  \left( \frac{64}{15} \zeta_3 + \frac{3334}{225} \zeta_2
                - \frac{191416183}{1620000} \right)
  + C_A^2  \left( - 62 \zeta_4 + \frac{44}{3} \zeta_3 - \frac{8213}{450} \zeta_2
                + \frac{122348527}{405000} \right) \,.
\label{eq:HiggsbulkNLO}
\end{align}
Combined with the singular prediction in the $z\to 1$ limit \cite{deFlorian:2004mp,Moult:2018jzp}, as well as the $\cO(\alpha_s^2)$ $\delta(1-z)$ term \cite{TZY:forthcoming}, this information enables us to extract the jet function constants. We find 
\begin{align}
j_2^q =&\ C_F n_f\left( \frac{9}{5} \zeta_2 + \frac{703847}{24000}\right) \nn
+ C_F C_A \left( - 76 \zeta_4 + 280 \zeta_3 + \frac{1063}{15} \zeta_2
                 - \frac{164883727}{324000}   \right) \nn\\
& + C_F^2\left( 152 \zeta_4 - 478 \zeta_3 - 106 \zeta_2 + \frac{3498505}{5184} \right)\,, \label{jq2} \\
j_2^g =&\ n_f^2\left( - \frac{8}{15} \zeta_2 + \frac{2344}{1125} \right) + 
  C_F n_f\left(4 \zeta_3 + \frac{14}{5} \zeta_2 - \frac{1528667}{108000} \right) \nn \\
& + C_A n_f\left(\frac{44}{5} \zeta_3 - \frac{127}{25} \zeta_2 + \frac{68111303}{1620000}   \right) 
+ C_A^2 \left( 76 \zeta_4- \frac{1054}{5} \zeta_3 - \frac{2159}{75}  \zeta_2 + \frac{133639871}{810000} \right) \,.
\label{jg2}
\end{align}
\end{widetext}
We have also checked this result by a direct calculation of the $n_f^2$ terms. Finally, in ref.~\cite{Korchemskytocome}, the idea of the sum rule presented in this section was extended to derive sum rules for $\int_0^1 dz\, z d\sigma/dz$, and $\int_0^1 dz\, (1-z) d\sigma/dz$. The additional weighting eliminates either the $\delta(z)$ or $\delta(1-z)$ term in the cross section, allowing the $\cO(\alpha_s^2)$ $\delta(z)$ term to be obtained independently of the $\cO(\alpha_s^2)$ $\delta(1-z)$ term. We have verified that these extended sum rules are satisfied to $\cO(\alpha_s^2)$ for all color channels, providing a stringent check of our jet function constants in Eq.~\eqref{jg2}, and emphasizing the interesting constraints on the EEC imposed by sum rules.

\section{NNLL Resummation in QCD}
\label{sec:NNLL}

With the two loop jet constants in hand, we are able to compute the all orders singular behavior of the EEC in the collinear limit to NNLL. The analytic solution of the renormalization group equations in QCD is complicated by the presence of the matrix structure, and the running coupling. We therefore solve the equation iteratively. Results to nine-loop order are provided in ancillary files for both $e^+e^-$ annihilation and gluonic decays of the Higgs. This order suffices for convergence down to $z=0.004$, and higher orders would be straightforward to obtain as well.  In App.~\ref{app}, we provide the timelike moments of the splitting functions that are necessary to perform the evolution, as well as the hard function coefficients for the two processes.

\begin{figure*}[t]
\centering
\subfloat[]{\label{fig:ee}
\includegraphics[width=0.48\textwidth]{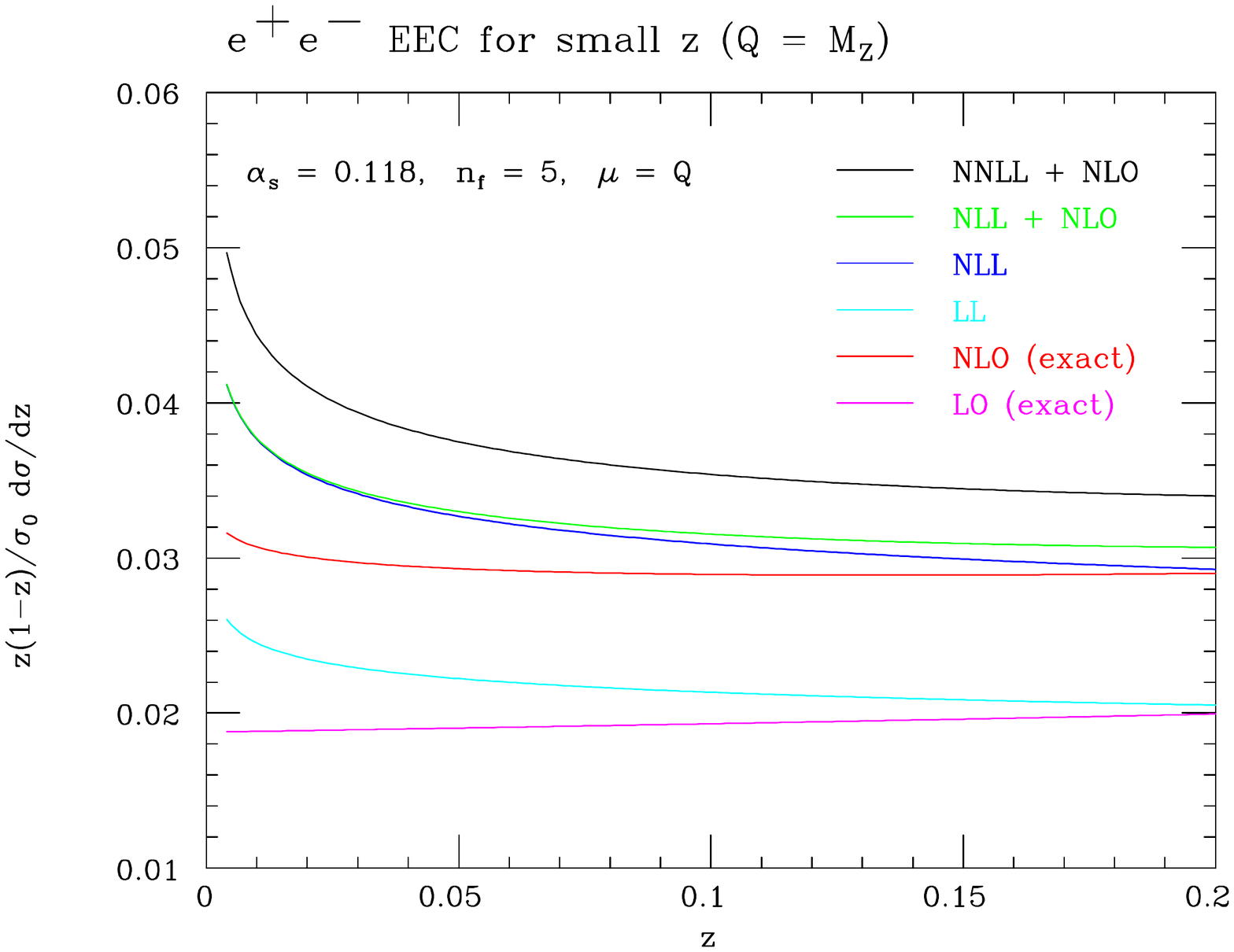}
}
\subfloat[]{\label{fig:higgs}
\includegraphics[width=0.48\textwidth]{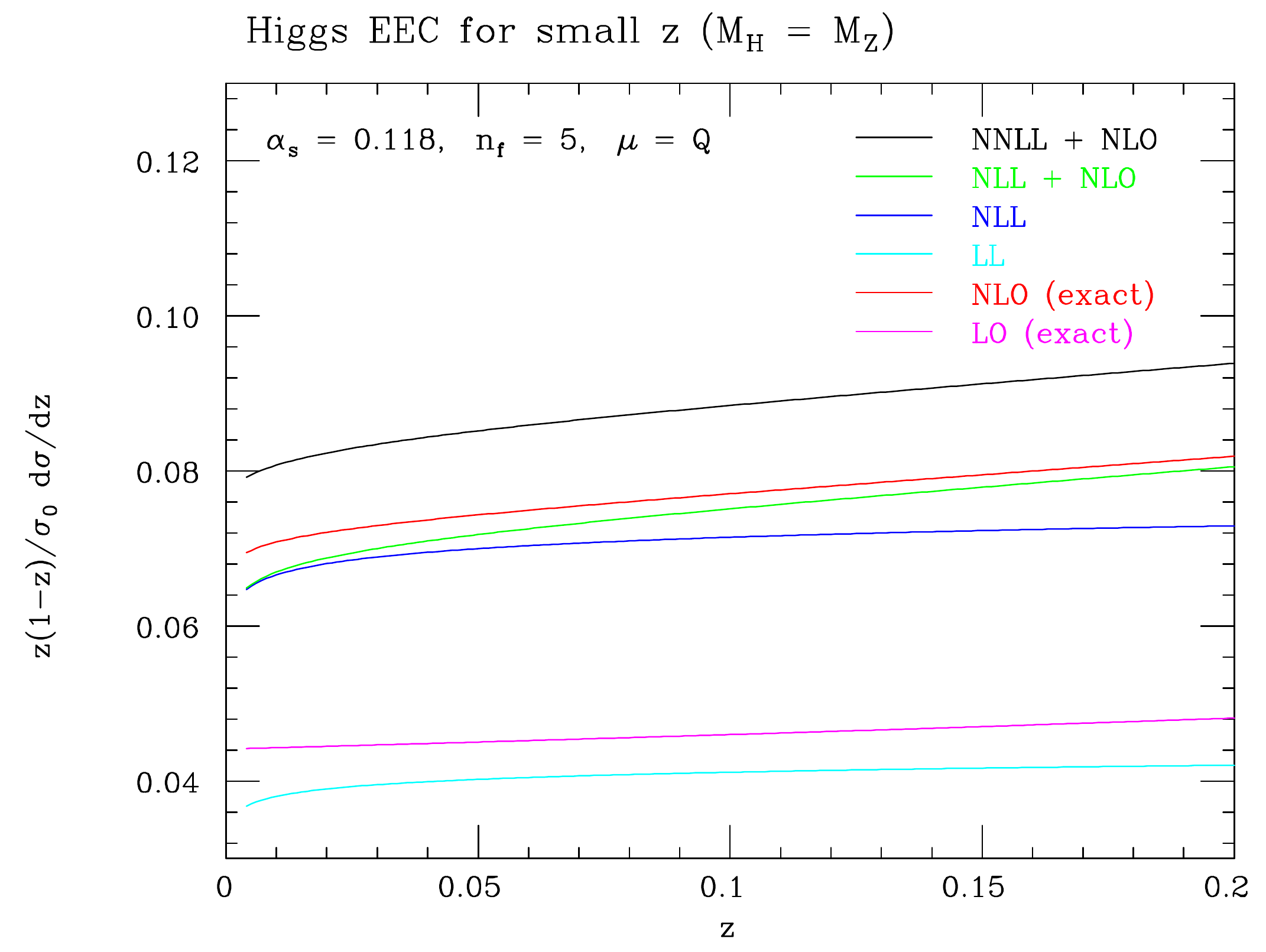}
}
\caption{Exact and resummed results for the EEC in the collinear limit for $e^+e^-$ annihilation in (a) and for Higgs decays to gluons in (b). Large perturbative corrections, driven in the $e^+e^-$ case partly by the $\beta$ function, are observed at each order.}\label{fig:fig1}
\end{figure*}

In \Fig{fig:fig1} we plot the resummed results in the $z\to 0$ limit for both $e^+e^-$ annihilation and Higgs decays to gluons at various logarithmic accuracies, for $\mu=Q$.  We match the NLL and NNLL resummations to the analytic NLO results \cite{Dixon:2018qgp,Luo:2019nig} by adding the resummed and NLO formulas and subtracting the overlapping $\alpha_s$ and $\alpha_s^2$ terms in the perturbative expansion of the resummed formula.  We take $\alpha_s(Q)=0.118$ and $n_f=5$, as appropriate for measurements at $Q=M_Z$.  To facilitate the comparison of quark and gluon sources, we have set the Higgs mass $M_H=M_Z$, and we do not include renormalization of the short-distance operator $H G_{\mu\nu} G^{\mu\nu}$. The higher order logarithmic corrections are large.  The right side of the plot shows that the large corrections extend out to moderately small angles, as was also observed in a fixed-angle NNLO computation for $e^+e^-$~\cite{DelDuca:2016csb}.

Note that we plot the EEC with a prefactor of $z(1-z)$.  In principle, our logarithmic resummation is insensitive to the factor of $(1-z)$, since it represents a subleading power correction.  However, comparing the expansion of the resummed formula with analytic fixed order results, we find that the LO and NLO power corrections are much smaller for $e^+e^-$ if we interpret the resummation as being for $z(1-z)/\sigma_0 \times d\sigma/dz$, rather than for $z/\sigma_0 \times d\sigma/dz$, so this is what we have done.  The small size of the power corrections resulting from this choice is visible on the right side of \Fig{fig:fig1}(a) (where the resummed terms are small) in the good agreement between the LL and LO (exact) curves, and between the NLL and NLO (exact) curves. The power corrections are larger in the Higgs case. It would be interesting to extend this comparison to NNLO~\cite{DelDuca:2016csb}.

In \Fig{fig:fig1} we observe quite different numerical behavior in the $z\to 0$ limit for the case of $e^+e^-$ annihilation and gluonic Higgs decays. This difference is due to the different collinear structure of the initiating hard partons, namely quarks in the case of $e^+e^-$ annihilation and gluons in the case of Higgs decays. To better understand this behavior, we recall that in a CFT the anomalous dimensions of twist-two operators are non-negative \cite{Ferrara:1974pt,Mack:1975je}. This guarantees that in a CFT, the differential cross section plotted as $z d\sigma/dz$ decreases as $z\to 0$. (See Eq.~\eqref{Neq4NNLLcumulant} for the form of the cumulant for a CFT.) In the case of QCD, there is a competition between $\beta$ function contributions and twist-two anomalous dimensions. The $\beta$ functions contributions drive $z d\sigma/dz$ larger as $z\to 0$, because the coupling is larger at smaller scales.  The twist-two anomalous dimensions, as in a CFT, drive $z d\sigma/dz$ smaller as $z\to 0$.  The competition plays out differently for quarks versus gluons.  For gluons the splitting anomalous dimensions win, leading to a suppression at small values of $z$, and comparatively ``wider" jets than for quarks, where the $\beta$ function contribution wins.  In other words, for the Higgs, the EEC behaves quite similarly to the case of a CFT, while for $e^+e^-$, the growth of the cross section as $z\to 0$ indicates a qualitatively different behavior than in a CFT. The balance between beta function contributions and anomalous dimensions is quite delicate, and as we will see in Sec.~\eqref{sec:landau}, in $\cN=1$ SYM we can exactly balance the two contributions at LL accuracy, so that there are in fact no leading logarithms as $z\to 0$!

This dependence on the source (or hard initiating parton) in the $z\to 0$ limit should be contrasted with the behavior in $z\to 1$ limit, where to LL accuracy we have \cite{Collins:1981uk,Moult:2018jzp}
\begin{align}
\label{eq:LLzto1}
\frac{1}{\sigma_0}  \frac{d\sigma(z)}{dz}  =  \frac{1}{8(1-z)}\int\limits_0^\infty db\, b J_0(b) e^{-\frac{1}{2}C_i \Gamma_{\text{cusp}} \ln^2 \left(\frac{e^{2\gamma_E} b^2}{4(1-z)} \right) }\,,
\end{align}
where $\Gamma_{\text{cusp}}$ is the cusp anomalous dimension \cite{Korchemsky:1987wg}, $J_0(b)$ is a Bessel function, and $C_i$ is the color Casimir, namely $C_i=C_F$ for $e^+e^-$ annihilation, and $C_i=C_A$ for Higgs decays to gluons. To this order, the only process dependence enters through the color Casimir, a property referred to as Casimir scaling, which is also observed for most jet substructure observables. We believe that the fact that the EEC is directly sensitive to the collinear structure of the initiating hard parton, beyond simply its color Casimir, makes it interesting as a jet substructure observable, and complementary to other such observables.

To understand the large corrections from LL to NLL to NNLL, we give the results through NNLO in the collinear limit, with $C_F = 4/3$, $C_A=3$, $n_f=5$, and $\mu = Q$ substituted in to simplify the expression. For the case of $e^+e^-$ annihilation we have
\begin{widetext}
\begin{align}
\label{eq:QCDas3analytic}
  &\frac{z}{\sigma_0} \frac{d\sigma^{e^+e^-}(z)}{dz} = \, 2a_s + a_s^2 \left( -\frac{173}{15} \ln z + \frac{16}{9} \zeta_3
- \frac{424}{27} \zeta_2 + \frac{638941}{6075} \right)
\\
&\, + a_s^3 \left[ \frac{20317}{450} \ln^2\!z 
+ \ln z \left(\frac{3704}{81} \zeta_3
         - \frac{343252}{1215} \zeta_2 - \frac{686702711}{1093500} \right) \right.\nn \\
 &\hspace{4cm}\left.+ \frac{352}{27} \zeta_5 + \frac{160}{9}  \zeta_2\zeta_3
   - \frac{8930}{81} \zeta_4 - \frac{633376}{405} \zeta_3
 - \frac{18994669}{36450} \zeta_2
               + \frac{745211486777}{131220000}
\right]+\mathcal{O}(\alpha_s^4)   \nn\\
&\hspace{0.9cm}=\, 2 a_s + a_s^2 (-11.5333 \ln z + 81.4809)
+a_s^3 \left( 45.1489 \ln^2 \!z - 1037.73 \ln z + 2871.36 \right) \,, \label{eq:QCDas3numerical}
\end{align}
and for the case of gluonic decays of the Higgs, we have
\begin{align}
\label{eq:Higgsas3analytic}
&\frac{z}{\sigma_0} \frac{d\sigma^{H}(z)}{dz} =\frac{47}{10} a_s +a_s^2 \left( \frac{2167}{150}\ln z -36 \zeta_3 +\frac{512}{5}\zeta_2 +\frac{2159543}{9000}  \right)  \\
&+a_s^3 \left[  -\frac{14117}{1125} \ln^2\!z + \ln z  \left( -\frac{28748}{135}\zeta_3 -\frac{321242}{2025}\zeta_2 +\frac{27672101}{18225} \right)\right. \nn \\
& \hspace{4cm}\left.+ 1296 \zeta_5 -\frac{86639}{27}\zeta_4  -\frac{4667179}{2025}\zeta_3+\frac{217606907}{40500}\zeta_2 +\frac{5406051434989}{437400000} \right] +\mathcal{O}(\alpha_s^4)\,,\nn \\
&=4.7 a_s +a_s^2(14.4467 \ln z+365.116) +a_s^3 (-12.5484 \ln^2\!z+1001.43 \ln z+ 16298.1) \,,  \label{eq:Higgsas3numerical}
\end{align}
\end{widetext}
where $a_s = \alpha_s(Q)/(4 \pi)$. The complete $C_F, C_A, n_f$ dependence can be found in the ancillary files. The ${\cal O}(\alpha_s^2)$ terms agree with the NLO fixed-angle result~\cite{Dixon:2018qgp}, also when the same analysis is applied to the Higgs case~\cite{Luo:2019nig}. Here we can clearly see the different signs for the logarithmic terms between the $e^+e^-$ and Higgs cases, explaining the behavior seen in \Fig{fig:fig1}.

The rapid growth of the perturbative coefficients is driven partly by the $\beta$ function, particularly for the case of $e^+e^-$, where the $\beta$ function drives the growth of the cross section as $z\to 0$. To see this, we can go to the Banks-Zaks fixed point \cite{Banks:1981nn}, letting $C_A=3$, $C_F=4/3$ and adjusting $n_f = 33/2 + {\cal O}(\alpha_s)$ in order to set $\beta_0=\beta_1=\beta_2=0$. 
We then find
\begin{align}
&\frac{z}{\sigma_0} \frac{d\sigma^{e^+e^-}(z)}{dz}
=2 a_s + a_s^2 ( 2.01111 \ln z - 2.22206 )  \\ 
&\hspace{1cm}+a_s^3( - 70.7058 \ln^2 \!z + 87.8276 \ln z - 490.324 )\,. \nn
\end{align}
We see that at the Banks-Zaks fixed-point there is a large reduction in the growth of the higher order perturbative corrections, although more than just the $\beta$ function is involved in the reduction of the $a_s^2 \ln^0 \!z$ term.  Also, for the Higgs case, where the logarithmic corrections are negative, we do not find that the Banks-Zaks values are smaller.  The poor convergence for QCD with five flavors motivates extending our results to N$^3$LL to obtain a more stable prediction.  One would also like to better understand qualitatively the dominant corrections at higher perturbative orders. One example could be to study the large $\beta_0$ limit which has previously been considered for non-singlet anomalous dimensions in QCD \cite{Gracey:1994nn,Gardi:2005yi}.

\section{\texorpdfstring{$\cN=1$}{N=1} SYM and Landau Poles}
\label{sec:landau}

To further illustrate the role of the $\beta$ function in the collinear limit, we consider pure $\cN=1$ SYM theory with an adjoint gluino. Results for this theory can be obtained from QCD by setting $C_F = C_A$, and $n_f = C_A$. (Such results are in the non-supersymmetric $\overline{\rm MS}$ scheme. They could be converted to the supersymmetric $\overline{\rm DR}$ scheme by a suitable redefinition of $\alpha_s$, but we won't do that here.)  In this case, one finds a fascinating cancellation due to the fact that $\sum_j \gamma_{jq}^{(0)}(3) = \sum_j \gamma_{jg}^{(0)}(3) = \beta_0 = 3C_A$. The anomalous dimensions and $\beta$ function therefore exactly cancel each other, and there is no leading logarithm.

Due to the simpler structure of this theory, we can write a closed form solution for the resummed cross section, which to NNLL reads, for $\mu=Q$,
\begin{align}
  \label{eq:Neqone}
   & \Sigma_{\rm NNLL}^{{\cal N}=1}(z) =c^S_1(\alpha_s)+c^S_2(\alpha_s) \ln z+c^S_3(\alpha_s) \frac{\ln z}{1+\beta_0 a_s \ln z}\nn \\
   &+c^S_4(\alpha_s) \ln[1+\beta_0 a_s \ln z]  \nn \\
   & +c^S_5(\alpha_s) \ln\left(1-2 C_A a_s \frac{\ln[1+\beta_0 a_s  \ln z] }{1+\beta_0 a_s \ln z} \right)\,.
\end{align}
Here the constants $c^S_i$ are functions of the coupling, and depend on the nature of the source, $S$. They can be found in App.~\ref{app} for a vector source coupled to quarks ($e^+e^-$) and scalar source coupled to gluons (Higgs). It would be interesting to explore the implications of $\cN=1$ supersymmetry for the constants, as has been done for conformal operators~\cite{Belitsky:1998gu}.  The last term in \eqn{eq:Neqone} comes from the form of the (logarithm of) the two-loop running coupling, with $\beta_1 = 6 C_A^2 = 2 C_A \beta_0$.  In QCD, the three-loop running coupling contributes at NNLL, but in $\cN=1$ SYM only two loops is required due to the leading-log cancellation mentioned above.

\begin{figure}[h]
  \centering \includegraphics[origin=c,width=0.48\textwidth]{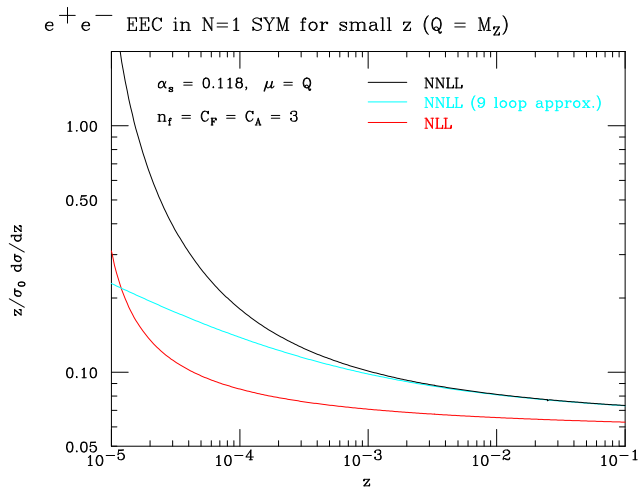}
  \caption{Resummed results for the EEC in $\cN=1$ SYM for an $e^+e^-$ source, using \eqn{eq:Neqone} at NNLL, and a simpler formula that resums the logarithms at NLL only.  We also plot the NNLL result using the same iterative approach used for QCD through nine loops.}
  \label{fig:fig2}
\end{figure}

In \Fig{fig:fig2} we plot the closed-form solution~\eqref{eq:Neqone} (NNLL), as well as an analogous solution at NLL, for the case of an $e^+e^-$ source.  (The Higgs source is qualitatively similar.)  The plot extends down to much smaller angles than the QCD plots in \Fig{fig:fig2}.  From the log-log plot it is clear that the result is far from a power law at these angles, where the coupling is varying rapidly.  It is still close to a power law for $z>0.004$, the range covered in the QCD plots. (Indeed the pure resummed QCD results are close to power law there too, because the QCD coupling is still not that large.)  We also provide the NNLL results in the same iterative nine-loop approximation we used for QCD, so that one can see how the approximation breaks down at smaller angles.

The closed form expression~\eqref{eq:Neqone} explicitly exhibits the Landau pole at $z \approx \exp[-1/(3 C_A a_s)] \approx 7\times 10^{-6}$ for $\alpha_s = 0.118$.  As shown in \Fig{fig:fig2}, the Landau pole has a positive residue. That is, in $\cN=1$ SYM theory the $\beta$ function dominates over the splitting anomalous dimension (for $e^+e^-$ or Higgs sources), starting at NLL, as was the case for $e^+e^-$ annihilation in QCD discussed earlier (see \Fig{fig:fig1}), starting at LL, although in that case, we did not obtain a closed form solution exhibiting the Landau pole. This feature highlights the important fact that if one is sufficiently far from the conformal limit that the $\beta$ function dominates over the splitting anomalous dimensions, then one can only compute the EEC perturbatively for values of $z$ greater than some minimal value, and the observable is not small in the $z\to0$ limit. In fact, it is so large that the sum rule~\eqref{eq:sumrule}, evaluated at finite coupling instead of order by order, does not converge at $z=0$.  It seems that in this case, some non-perturbative input is required, and it would be nice to know if the sum rule could provide constraints. The single-logarithmic nature of the small-angle EEC is quite different than a Sudakov limit in which the double logarithms in \eqn{eq:LLzto1} provide a strong exponential suppression as one approaches the infrared. 

\section{\texorpdfstring{${\cal N}=4$}{N=4} SYM and Reciprocity}
\label{sec:N4}

In this section, we apply our framework to ${\cal N}=4$ SYM theory, which is a CFT, leading to a simple behavior in the collinear limit based on a spacelike OPE~\cite{Hofman:2008ar,DavidPetrSashaTalks,DSDtocome,Korchemskytocome}. In addition to highlighting the different behavior in a CFT, ${\cal N}=4$ SYM theory is particularly interesting because the anomalous dimension that governs the singular behavior can be determined to high orders in the weak coupling expansion, or even at finite coupling from integrability. Therefore the collinear limit can be studied at a level that is unachievable in QCD. Furthermore, the study of the singular limits provides data to potentially enable a bootstrap of the complete result for the EEC.

In ${\cal N}=4$ SYM, supersymmetry implies that $\sum_j P_{j\phi}(y) = \sum_j P_{j\lambda}(y) = \sum_j P_{jg}(y) = P_{T, {\text{uni.}}}(y)$, where $j$ is summed over the scalar $\phi$, fermion $\lambda$, and gluon in the ${\cal N}=4$ super-multiplet, and $P_{T, \text{uni.}}$ is a universal timelike splitting kernel \cite{Kotikov:2004er}. Therefore, the splitting matrix reduces to a scalar, significantly simplifying the analysis of the evolution equations. Furthermore, the result is independent of the source for any operator in the stress-tensor multiplet \cite{Korchemsky:2015ssa,Belitsky:2014zha}.

More interestingly, since the coupling does not run in a CFT, the only scale in the problem is $z Q^2$. One can then make a power law ansatz for the jet function
\begin{align}
  \label{eq:Jneqfour}
  J(z Q^2,\mu) = C_J(\alpha_s) \left(\frac{z Q^2}{\mu^2} \right)^{\gamma_J^{{\cal N}=4}(\alpha_s)} \,,
\end{align}
where the anomalous dimension $\gamma^{{\cal N}=4}(\alpha_s)$ can be determined by substituting into the jet function evolution equation~\eqref{eq:jetRG}. Explicitly, using the definition~\eqref{eq:gammaPTrelation}, we find\footnote{We thank Simon Caron-Huot and Gregory Korchemsky for describing a version of this argument to us, motivated by our preliminary timelike results and the spacelike results of~\cite{DavidPetrSashaTalks,DSDtocome}, and pointing out the important connection to reciprocity.}
\begin{align}
  \label{eq:gammaNeqfour}
2 \gamma_J^{{\cal N}=4}(\alpha_s) = &\, - 2 \int_0^1 dy\, y^{2 + 2 \gamma_J^{{\cal N}=4}(\alpha_s)} P_{T, \text{uni.}}(x,\alpha_s)
\nn\\
=&\, 2 \gamma_T^{{\cal N}=4}(1 + 2 \gamma_J^{{\cal N}=4}, \alpha_s) \,,
\end{align}
where $\gamma_T^{{\cal N}=4}(N,\alpha_s)$ is the Mellin $N+2$ moment of the universal splitting kernel $P_{T, \text{uni.}}(x, \alpha_s)$. Note that in the $\cN=4$ case we use a shifted argument, since performing the sum $\sum_j \gamma_{j\phi}(N) = \sum_j \gamma_{j\lambda}(N) = \sum_j \gamma_{jg}(N) = \gamma_{T, \rm uni.}(N-2)$ shifts the argument by two units in Mellin space. Therefore, for the scalar $\cN=4$ universal anomalous dimension, although it is evaluated at $N=1$, we will still refer to it as the twist two spin three anomalous dimension. 

When the power-law behavior of the jet function~\eqref{eq:Jneqfour}
in $\cN=4$ SYM is inserted into the factorization formula~\eqref{eq:fact},
the $z$ dependence can be factored out of the integral.
We therefore find that in $\cN=4$ SYM, the $z\to0$ asymptotics can be
written as a simple power law
\begin{equation}
\Sigma(z) = \frac{1}{2} \, C(\alpha_s) \, z^{\gamma_J^{{\cal N}=4}(\alpha_s)} \,,
\label{Neq4NNLLcumulant}
\end{equation}
as is expected for the scaling behavior of a CFT.  This simple power law should be contrasted with the more complicated behavior in a non-CFT, for example \eqn{eq:Neqone}. The $\cN=4$ result can also be written as a power series in $\ln z$, which is given at NNLL in App.~\ref{app}.  

To further simplify the quantity $\gamma_J^{{\cal N}=4}$ appearing in \eqn{Neq4NNLLcumulant}, we can combine Eq.~\eqref{eq:gammaNeqfour} with the reciprocity relation between timelike and spacelike anomalous dimensions \cite{Mueller:1983js,Dokshitzer:2005bf,Marchesini:2006ax,Basso:2006nk,Dokshitzer:2006nm},\footnote{Note that the term reciprocity is sometimes used to refer to the fact that the anomalous dimensions of a CFT are functions of the conformal spin \cite{Dokshitzer:2005bf,Basso:2006nk} which was proven to all orders in perturbation theory~\cite{Alday:2015eya}. Here we use reciprocity in a stronger sense, namely that when expressed in terms of the conformal spin, both the spacelike and timelike anomalous dimensions can be written in terms of the same universal function \cite{Dokshitzer:1995ev,Dokshitzer:2005bf}, leading to the functional relation between the spacelike and timelike anomalous dimensions in Eq.~\eqref{eq:reciprocity} \cite{Mueller:1983js,Dokshitzer:2005bf,Marchesini:2006ax,Basso:2006nk,Dokshitzer:2006nm}. To our knowledge, there does not exist an all orders proof of this relation, although, as mentioned in the text, the equivalence of the results of \cite{DavidPetrSashaTalks,DSDtocome,Korchemskytocome} with those presented here allows it to be proven for one moment, $N=1$ in $\cN=4$ parlance.}
\begin{align}\label{eq:reciprocity}
2 \gamma_S^{{\cal N}=4}(N , \alpha_s)=  2 \gamma_T^{{\cal N}=4}(N + 2 \gamma_S^{{\cal N}=4}, \alpha_s), 
\end{align}
to find that
\begin{align}
 \gamma_J^{{\cal N}=4}(\alpha_s) = \gamma_S^{{\cal N}=4}(1,\alpha_s).
 \end{align}
In other words, the scaling evolution of the jet function is governed by the universal anomalous dimension of the spacelike twist two spin three operator!  Furthermore, as mentioned above, in a CFT the anomalous dimensions of spacelike twist-two operators are positive, guaranteeing that the resummed result for the differential cross section is integrable in the $z\to 0$ limit.  The spacelike twist-two anomalous dimensions are particularly convenient since they are anomalous dimensions of local operators.  In $\cN=4$ SYM, they can be computed up to a remarkable 7 loops \cite{Kotikov:2004er,Kotikov:2006ts,Kotikov:2002ab,Kotikov:2003fb,Bajnok:2008qj,Kotikov:2007cy,Velizhanin:2014zla,Velizhanin:2014fua,Lukowski:2009ce,Velizhanin:2013vla,Marboe:2014sya,Marboe:2016igj}, and numerically at finite coupling using the quantum spectral curve \cite{Gromov:2013pga,Gromov:2014caa,Gromov:2014bva,Gromov:2015wca}.

It is quite remarkable that the timelike dynamics of a jet can be described by the anomalous dimension of local operators, at least in a CFT. This was first observed in ref.~\cite{Hofman:2008ar} and has been studied in refs.~\cite{DavidPetrSashaTalks,DSDtocome} using the light-ray operator formalism~\cite{Kravchuk:2018htv,Kologlu:2019bco}, and also in ref.~\cite{Korchemskytocome} using a Mellin-based approach. Here we have shown how the reciprocity relation provides a connection between this framework and the more standard timelike splitting picture used to study the dynamics of jets in QCD. Alternatively, the equivalence of the results of \cite{DavidPetrSashaTalks,DSDtocome,Korchemskytocome}, which are naturally expressed in terms of spacelike data, and our results, which are naturally expressed in terms of timelike data, allow for a proof of the reciprocity relation, Eq.~\eqref{eq:reciprocity}, for one value of the Mellin moment. We believe that further studies of the relationship between the spacelike and timelike approaches could provide a better understanding of reciprocity relations.

The constant $C(\alpha_s)$ in~\eqn{Neq4NNLLcumulant} is given by 
\begin{equation}
C(\alpha_s)\  =\ 
1 - \frac{C_A \alpha_s}{\pi}
  + \biggl( \frac{11}{4} \zeta_4 - 3 \zeta_2 + 7 \biggr)
  \biggl(\frac{C_A \alpha_s}{\pi}\biggr)^2
  + {\cal O}(\alpha_s^3), \label{Cresult}
\end{equation}
and the spacelike anomalous dimension is given by
\begin{align}
  \label{eq:eta}
  \gamma_S^{{\cal N}=4}(1,\alpha_s) =& \ 
\frac{C_A \alpha_s}{\pi}
 + \biggl( - \frac{\zeta_3}{2} + \zeta_2 - 2 \biggr)
     \biggl(\frac{C_A \alpha_s}{\pi}\biggr)^2
\nn\\
                  &\hspace{-2.0cm}
+ \biggl( \frac{3}{2} \zeta_5 + \frac{3}{8} \zeta_4
         - \frac{3}{2} \zeta_3 - 4 \zeta_2 + 8 \biggr) 
    \biggl(\frac{C_A \alpha_s}{\pi}\biggr)^3 \nn \\
&\hspace{-2cm}+\left(-\frac{69}{16}\zeta_7 + \frac{1}{2}\zeta_2 \zeta_5 - \frac{5}{16}\zeta_3 \zeta_4 + \frac{9}{4}\zeta_3^2 - \frac{107}{32}\zeta_6 + 8\zeta_5\right. \nn \\
&\hspace{-1.4cm} \left.-  \frac{13}{2}\zeta_2 \zeta_3 - \frac{23}{8}\zeta_4 + 7\zeta_3 + 24\zeta_2 - 40 \right)  \biggl(\frac{C_A \alpha_s}{\pi}\biggr)^4 \nn \\
&\hspace{-2cm}+ \Ord(\alpha_s^5) \,.
\end{align}
The expression for the spacelike anomalous dimension is non-standard, since it has been continued to odd $N$ \cite{Kotikov:2005gr}.  The result~\eqref{eq:eta} agrees with an independent computation~\cite{Korchemskytocome}.

The ${\cal O}(\alpha_s^2)$ term in $C(\alpha_s)$ was extracted from the sum
rule~\eqref{eq:sumrule}, using an analysis of the
back-to-back limit~\cite{TZY:forthcoming,Korchemskytocome},
and the bulk integral computed from the NLO result~\cite{Belitsky:2013ofa},
\begin{align}
& \frac{1}{\sigma_0} \int_0^1 dz \frac{d\sigma^{\cN=4,{\rm bulk}}}{dz} = 
\frac{C_A\alpha_s}{2\pi} ( \zeta_2 + 1 ) \nn\\
&
+ \left(\frac{C_A\alpha_s}{2\pi}\right)^2
 \Bigl( - \frac{31}{2} \zeta_4 + 6 \,\zeta_2 - 14 \Bigr)
+ {\cal O}(\alpha_s^3)\,.
\label{eq:Neq4bulk}
\end{align}
Note that in $\cN=4$ SYM, corrections to the total cross section vanish to all orders for the standard source because it is a protected operator. Also, unlike in QCD, it is not necessary to distinguish the jet and hard contributions to the $\delta(z)$ term, because the coupling does not run and so it is the same at the natural scales for both functions, $\sqrt{z}Q$ and $Q$. Differentiating \eqn{Neq4NNLLcumulant} with respect to $z$ and expanding in $\alpha_s$, we find complete agreement with all the $\alpha_s^3$ terms appearing in the $\chi\to 0$ limit of the recent NNLO fixed-angle result~\cite{Henn:2019gkr}.

Recently it has become possible to use an OPE computation~\cite{Eden:2012rr} to determine $C(\alpha_s)$ to ${\cal O}(\alpha_s^3)$, i.e.~N$^3$LL, and the back-to-back limit is also understood at this order~\cite{Korchemskytocome}.  The sum rule~\eqref{eq:sumrule} then predicts the next term in \eqn{eq:Neq4bulk}, which can be computed~\cite{Korchemskytocome} using the results of ref.~\cite{Henn:2019gkr}.  It would be interesting to see whether the normalization coefficient, $C(\alpha_s)$ could be extracted to even higher orders, or even exactly, using integrability.

Through at least three loops, anomalous dimensions of twist two operators obey a principle of maximal transcendentality~\cite{Kotikov:2001sc,Kotikov:2002ab,Kotikov:2004er,Kotikov:2007cy}:  The $\cN=4$ results are harmonic sums with a uniform transcendental weight, $2L-1$ in Mellin space at $L$ loops, and they can be extracted from the QCD results by setting $C_F \to C_A$ and keeping only the leading transcendental terms.  This principle does not work for the EEC at fixed angles, i.e.~generic $z$; the leading transcendental functions of $z$ have different rational prefactors.  In the back-to-back limit, $z\to1$, large spin operators dominate, the $\cN=4$ SYM EEC has a uniform weight, and the principle of maximal transcendentality holds.  In the collinear limit, $z\to0$, an operator of fixed spin dominates, and the harmonic sums evaluate to rational numbers that do not convey the weight information anymore.  Nevertheless, by comparing the $\cN=4$, $\cN=1$ and QCD results for the EEC as $z\to0$, and counting $\ln z$ as weight 1, we see that the terms of maximal transcendental weight $2L-1$ are equal.  This property is ultimately inherited from the fixed-spin spacelike (or timelike) anomalous dimensions.

We can also assess the other individual contributions to the sum rule~\eqref{eq:sumrule} that are of leading transcendentality, in this case weight $2L$. We first observe that the leading transcendental terms in the $\cN=4$ bulk integral~\eqref{eq:Neq4bulk} agree with those in the QCD bulk expressions~\eqref{eq:QCDbulkNLO} and \eqref{eq:HiggsbulkNLO}, after setting $C_F\to C_A$.  The $\delta(z)$ coefficients, which were used to fix $j_2^q$ and $j_2^g$, also have this property. The corrections to the total cross section vanish in $\cN=4$ SYM, but not in QCD; however, the QCD corrections have subleading transcendentality.  In other words, each of the four individual contributions to the sum rule~\eqref{eq:sumrule} appears to separately obey a leading transcendentality principle, although only the $\delta(1-z)$ term is of homogeneous weight in $\cN=4$ SYM. 

We conclude this section by discussing to what extent reciprocity can be used to organize the timelike evolution in a non-conformal field theory. While the relation~\cite{Basso:2006nk}
\begin{equation}
\label{eq:BKrelation}
2\gamma_S(N) = 2\gamma_T(N+2\gamma_S(N)),
\end{equation}
is expected to hold in a non-conformal theory, the property that only the $N=3$ moment contributes to the small-angle EEC will not persist away from the conformal limit.

Consider for simplicity the case of pure Yang-Mills theory, e.g.~set $n_f=0$ for the gluonic source of Higgs decay.  We make an ansatz for the evolved gluon jet function that incorporates the running coupling,
\begin{align}\label{eq:ansatz_YM}
&J_g\left(\frac{\mu^2}{zQ^2}, \alpha_s(\mu) \right) \\
&\hspace{0.7cm} =\ C_J (\alpha_s(\sqrt{z}Q)) \exp \Biggl[
 - \! \!\!\int\limits_{\alpha_s(\sqrt{z}Q)}^{\alpha_s(\mu)}   \!\!\! d\bar \alpha_s \frac{\gamma^{\text{YM}}_J(\bar \alpha_s, z)}{\beta(\bar \alpha_s)} \Biggr] \,, \nn
\end{align}
with $\beta(\alpha_s) \equiv d\alpha_s(\mu)/d\ln\mu^2$,
in terms of an effective anomalous dimension $\gamma^{\text{YM}}_J(\alpha_s,z)$. Repeating the derivation given in this section for $\cN=4$ SYM, one finds that to NLL in $\ln z$ we have the relation
\begin{align}
2\gamma_J^{\text{YM}}=2\gamma^{\text{YM}}_{T}\left( 3+\frac{2\gamma_J^{\text{YM}}}{1+\frac{\alpha_s(Q)}{4\pi} \beta_0 \ln z   }    \right)\,.
\end{align}
Expanding in terms of $\beta_0\ll 1$, using the reciprocity relation of \eqn{eq:BKrelation}, and keeping only the terms to NLL, one finds
\begin{align}
\label{eq:pureYMNLL}
\gamma_J^{\text{YM}}&=\gamma^{\text{YM}}_S(3)  \\
&- \gamma^{\text{YM}}_S(N) \partial_N \gamma^{\text{YM}}_S(N)\Big|_{N=3}
\frac{\alpha_s(Q)}{4\pi} 2 \beta_0 \ln z +\cdots\,. \nn
\end{align}
Therefore, in a non-CFT, one no longer needs just $\gamma^{\text{YM}}_S(3)$, but also Mellin space derivatives around this point with coefficients proportional to the $\beta$ function. We emphasize that $\gamma_J^{\text{YM}}$ is the effective anomalous dimension defined by the ansatz~\eqref{eq:ansatz_YM}, which is why it has explicit $z$ dependence.

It would be helpful to understand \eqn{eq:pureYMNLL} from the perspective of a weakly broken conformal field theory, as well as to extend such a relation to the multi-flavor case. However, since in QCD the $\beta$ function and the twist-two anomalous dimensions are of the same order, this organization becomes increasingly complicated at higher orders (This was clearly illustrated in Sec.~\ref{sec:landau} where for the case of $\cN=1$ SYM the $\beta$ function exactly cancelled the running from the twist-two anomalous dimension at LL.). Another complication is that the couplings in the jet function and the hard function are naturally evaluated at different scales, namely $\alpha_s(\sqrt{z}Q)$ and $\alpha_s(Q)$, and it would be nice to explore how this arises from the spacelike perspective. We leave these directions to future work.

\section{Conclusions}
\label{sec:conclusion}

In this paper we have presented a factorization formula which describes the collinear limit $\chi\to 0$ of the EEC observable. This formula applies in a conformal or asymptotically free QFT, and is formulated in terms of the timelike data of the theory. For QCD and $\cN=1$ SYM, we computed the EEC to NNLL, extending the previously known jet calculus resummation at LL. In the particular case of a CFT, which here we took as $\cN=4$ SYM, we have shown how spacelike-timelike reciprocity allows the result to be written as a single power law with the spacelike $N=3$ moment, providing a connection with the approach of ref.~\cite{Hofman:2008ar}. We have also emphasized the importance of the sum rule in \eqn{eq:sumrule}, which allows the singular behavior in the $\chi\to 0$ and $\chi\to 1$ limits to be related to information in the bulk region of the EEC distribution.

There are a number of directions that would be interesting to pursue. First, for phenomenological applications, due to the large corrections observed at NNLL in QCD, it would be helpful to perform the resummation at N$^3$LL. This would allow the EEC to be described by N$^3$LL resummation of large logarithms at both $z\to 0$ and $z\to 1$ endpoints, combined with NNLO fixed order results in the bulk of the distribution.  One of the ingredients for resumming the $z\to0$ limit at N$^3$LL is the set of $N=3$ values of the N$^3$LO twist-two timelike anomalous dimensions, which should be obtainable from the spacelike ones using reciprocity. At present, the nonsinglet N$^3$LL spacelike anomalous dimensions are available for arbitrary Mellin moment in the large $N_c$ limit, and approximately for the subleading-in-$N_c$ terms~\cite{Moch:2017uml}.  A few moments of the singlet anomalous dimensions are available~\cite{Vogt:2018miu}, which might already allow for an approximate determination.  It will also be necessary to compute the hard functions at this order.  The three-loop jet functions may then be extractable using the sum rule for $\int_0^1 dz\, (1-z) d\sigma/dz$ \cite{Korchemskytocome}, if the Higgs EEC can be computed numerically at NNLO for generic angles.  Finally, in order to use such N$^3$LL results in a precision extraction of the strong coupling, $\alpha_s$, a good understanding of the non-perturbative corrections in the collinear limit will be required.


On the more formal side, it would be beneficial to explore to what extent reciprocity can shed light on the EEC in QCD, including both the effect of the running coupling, and multiple flavors. Reciprocity has also been observed at higher twist \cite{Beccaria:2007bb,Macorini:2010px}, and it would be interesting to extend our timelike factorization formula to higher powers in the $z$ expansion, and to understand the role that reciprocity plays at subleading powers. A better understanding might enable timelike dynamics to be related to local operators, which could then potentially allow them to be computed non-perturbatively on the lattice. It would also be interesting to better understand the relation between the timelike factorization approach presented in this paper, and the recent approaches of refs.~\cite{Belin:2019mnx,DavidPetrSashaTalks,DSDtocome,Korchemskytocome}.

Finally, our factorization formula, with the same jet functions but modified hard functions, also applies to small-angle energy correlations that can be measured at a hadron collider such as the LHC. Observables similar to the EEC are commonly used in jet substructure \cite{Larkoski:2013eya,Larkoski:2014gra,Larkoski:2015kga,Moult:2016cvt,Komiske:2017aww} (for a review see ref.~\cite{Larkoski:2017jix}). Note that the EEC, unlike typical event classifiers, produces a distribution of values even for a single event. In this context, the EEC provides an interesting example of a single logarithmic jet substructure observable that is directly sensitive to the collinear structure of jets, and is naturally insensitive to soft radiation. \Fig{fig:fig1} exhibits the different behavior of the EEC for quark and gluon jets. We therefore believe that the theoretical simplicity of the EEC in the collinear limit, and its relation to well known field-theoretic quantities, will enable further advances in our understanding of the substructure of jets.

\section{Acknowledgments}

We thank Andrei Belitsky, Simon Caron-Huot, Vittorio Del Duca, Claude Duhr, David Simmons-Duffin, Gregory Korchemsky, Juan Maldacena, and Tong-Zhi Yang for useful discussions, David Simmons-Duffin and Gregory Korchemsky for communicating their results for $\cN=4$ SYM to us prior to publication, and Gregory Korchemsky for helpful comments on the manuscript. L.D.~thanks Humboldt University Berlin and the University of Freiburg for hospitality while this project was completed. L.D.~is supported by the Office of High Energy Physics of the U.S.~DOE under Contract No.~DE-AC02-76SF00515 and by a Humboldt Research Award. I.M.~is supported by the Office of High Energy Physics of the U.S.~DOE under Contract No.~DE-AC02-05CH11231. H.X.Z.~is supported in part by the Fundamental Research Funds for the Central Universities under Contract No.~107201*172210191.


\appendix

\begin{widetext}

\section{Additional Perturbative Data}\label{app}

In this Appendix, we collect several additional results related to the perturbative behavior of the EEC in the collinear limit for $\cN=4$, $\cN=1$ SYM and QCD.

\subsection*{\texorpdfstring{$\cN=4$}{N=4} SYM:}

While the power law form of~\eqref{Neq4NNLLcumulant} is natural from the perspective of a CFT, for comparison with our results in QCD and $\cN=1$ SYM, it is interesting to also write the $\cN=4$ SYM result as a power series in $\ln z$. We find
\begin{align}
\frac{z}{\sigma_0} \frac{d\sigma}{dz}
&=\sum\limits_{L=1}^{\infty}\left( \frac{C_A \alpha_s}{\pi}  \right)^L  \frac{\ln^{L-1}z}{2(L-1)!} \nn \\
&-\sum\limits_{L=2}^{\infty}\left( \frac{C_A \alpha_s}{\pi}  \right)^L  \frac{\ln^{L-2}z}{2^2(L-2)!} \left[ (L-1)(\zeta_3-2\zeta_2)+2(2L-1)  \right] \nn \\
&-\sum\limits_{L=3}^{\infty}\left( \frac{C_A \alpha_s}{\pi}  \right)^L  \frac{\ln^{L-3}z}{2^4(L-3)!} \Big[(L-2)(L-3)(\zeta_3^2-4\zeta_2\zeta_3)+12(L-2)\zeta_5 +(10L^2-47L+76)\zeta_4 \nn \\ 
&\hspace{4.5cm}+8(L-2)(L-4)\zeta_3  
-8(2L^2-5L+5)\zeta_2+8(2L^2-1) \Big]\nn \\
&+\cO(\alpha_s^L \, \ln^{L-4} z)\,.
\end{align} 
Unlike the result for the non-conformal $\cN=1$ SYM theory in~\eqref{eq:Neqone}, the $\cN=4$ SYM result is a pure power series in $\ln z$, and does not involve $1/(1+\beta_0 a_s \ln z)$ terms which give rise to the Landau pole. In $\cN=4$ SYM, this series seems convergent for all values of $z$.

\subsection*{\texorpdfstring{$\cN=1$}{N=1} SYM:}

In the text we presented the form of the $\cN=1$ SYM result to NNLL as
\begin{align}
\label{eq:SigNeq1}
   & \Sigma_{\rm NNLL}^{{\cal N}=1}(z) =c_1^S + c_2^S \ln z + c_3^S \frac{\ln z}{1+\beta_0 a_s \ln z}
   +c_4^S \ln[1+\beta_0 a_s \ln z]  
    +c_5^S \ln\left(1-2 C_A a_s \frac{\ln[1+\beta_0 a_s  \ln z] }{1+\beta_0 a_s \ln z} \right)\,,
\end{align}
where $\beta_0=3C_A$ and the coefficients $c_i^S$ depend on the source.  Here we collect the coefficients $c_i^\gamma(\alpha_s)$ for an $e^+e^-$ source and $c_i^H$ for a Higgs source, in the $\overline{\rm MS}$ scheme. We find

\begin{align}
c_1^H&= \frac{1}{2}+\frac{69}{8} a
+a^2 \left( 22\zeta_4-66 \zeta_3-\frac{95}{3}\zeta_2+\frac{81949}{432} \right)  \,, \nn \\
c_1^\gamma&=\frac{1}{2}+\frac{13}{24} a +a^2 \left( 22\zeta_4-44 \zeta_3+\frac{22}{9}\zeta_2+\frac{2911}{162}  \right)\,, \nn \\
c_2^H&=\frac{3}{2}a +a^2\left(  -4\zeta_3 +\frac{3163}{72}\right) +a^3\left( \frac{16}{3}\zeta_3^2+24\zeta_5+16\zeta_2\zeta_3 -72 \zeta_4 -\frac{5656}{27}\zeta_3 -\frac{797}{6}\zeta_2 +\frac{1071895}{972} \right)  \,, \nn \\
c_2^\gamma&=  \frac{3}{2}a +a^2 \left(  -4\zeta_3+\frac{1417}{72} \right) +a^3\left(  \frac{16}{3}\zeta_3^2 +24\zeta_5+16 \zeta_2\zeta_3 -72\zeta_4-\frac{2128}{27}\zeta_3 -\frac{61}{2}\zeta_2 +\frac{1136527}{3888} \right)\,, \nn \\
c_3^\gamma&=c_3^H=a^3\left(  -\frac{16}{3}\zeta_3^2 +24 \zeta_5+16\zeta_2\zeta_3 -342 \zeta_4 +\frac{6097}{27}\zeta_3 +\frac{1243}{6} \zeta_2 -\frac{406067}{1944} \right)  \,, \nn \\
c_4^H&=a \left( 4\zeta_2-\frac{11}{3}  \right)
+ a^2 \left( 4\zeta_2-\frac{11}{3} \right)
      \left( - \frac{8}{3} \zeta_3 + \frac{3163}{108} \right) \,, \nn \\ 
c_4^\gamma&= a \left( 4\zeta_2-\frac{11}{3}  \right)
+ a^2 \left( 4\zeta_2-\frac{11}{3} \right)
      \left( - \frac{8}{3} \zeta_3 + \frac{1417}{108} \right) \,, \nn \\ 
c_5^H&= - a \left( 4\zeta_2-\frac{11}{3} \right)^2
- a^2 \left( 4\zeta_2-\frac{11}{3} \right)
      \left( - \frac{8}{3} \zeta_3 + \frac{3163}{108} \right) \,, \nn \\ 
c_5^\gamma&= - a \left( 4\zeta_2-\frac{11}{3} \right)^2
- a^2 \left( 4\zeta_2-\frac{11}{3} \right)
      \left( - \frac{8}{3} \zeta_3 + \frac{1417}{108} \right) \,, 
\end{align}
where $a \equiv C_A a_s = C_A \alpha_s/(4\pi)$.
From these results, we can clearly see that the leading transcendental pieces are equal for the two sources, and they are also equal to the leading transcendental pieces in $\cN=4$ SYM. (Note that one cannot drop all the $\beta$ function terms in \eqn{eq:SigNeq1} in checking this statement.) In $\cN=4$ SYM, the result is independent of the source, as long as it is in the stress energy supermultiplet \cite{Korchemsky:2015ssa,Belitsky:2014zha}; however, in $\cN=1$ SYM, this is no longer the case. It would be interesting to better understand the differences.

\subsection*{QCD:}

To iteratively solve the evolution equation for the jet function in Eq.~\eqref{eq:jetRG}, we require the $N=3$ moments of the timelike splitting functions, as well as certain logarithmic moments of the splitting functions, which occur when the equation is iterated to higher order. For convenience, in this appendix we collect all moments required to achieve NNLL accuracy, as well as the constants in the relevant hard functions.

We expand the timelike splitting functions perturbatively as
\begin{align}
P_{ij}(x)=\sum\limits_{L=0}^\infty
\left( \frac{\alpha_s}{4\pi} \right)^{L+1} P^{(L)}_{ij}(x)\,, 
\end{align}
and we denote the $N=3$ moment, which is relevant for the evolution of the EEC, by 
\begin{align}
\gamma_{T,ij}^{(L)} = - \int\limits_0^1 dx \, x^2 \, P^{(L)}_{ij}(x) \,.
\end{align}
To NNLL, we need the $N=3$ moment at LO, NLO and NNLO, which can be obtained from refs.~\cite{Mitov:2006wy,Mitov:2006ic,Moch:2007tx,Almasy:2011eq}. (Note that we include the pure singlet term in the $qq$ element.)  At LO, we have
\begin{align}
\gamma_{T,qq}^{(0)} &= \frac{25}{6} \, C_F\,, \qquad
\gamma_{T,gq}^{(0)}= -\frac{7}{6} \, C_F \,, \qquad
\gamma_{T,qg}^{(0)} = -\frac{7}{15} \, n_f \,, \qquad
\gamma_{T,gg}^{(0)} = \frac{14}{5}\, C_A + \frac{2}{3} \, n_f \,.
\end{align}
At NLO, we have
\begin{align}
\gamma_{T,qq}^{(1)}&=
\left( - 16 \zeta_3+ 24 \zeta_2-\frac{1693}{48} \right) C_F^2 
  + \left( 8\zeta_3 - \frac{86}{3} \zeta_2  +\frac{459}{8} \right) C_A C_F
  - \frac{5453}{1800} \, C_F n_f \,, \nn \\
\gamma_{T,gq}^{(1)} &= \left(\frac{28}{3}\zeta_2-\frac{2977}{432}  \right) C_F^2
+ \left(- \frac{14}{3}\zeta_2 -\frac{39451}{5400} \right) C_A C_F  \,, \nn \\
\gamma_{T,qg}^{(1)} &= \left(\frac{28}{15}\zeta_2 +\frac{619}{2700} \right) C_A n_f
- \frac{833}{216} \, C_F n_f - \frac{4}{25} \, n_f^2 \,, \nn \\
\gamma_{T,gg}^{(1)} &=
\left( - 8\zeta_3+ \frac{52}{15}\zeta_2+\frac{2158}{675}\right) C_A^2
+ \left(- \frac{16}{3}\zeta_2 +\frac{3803}{1350} \right) C_A n_f 
+ \frac{12839}{5400} \, C_F n_f \,.
\end{align}
At NNLO, we have
\begin{align}
\gamma_{T,qq}^{(2)}&=
\left(112\zeta_5+48\zeta_2\zeta_3- \frac{2083}{3} \zeta_4+ \frac{16153}{18} \zeta_3- \frac{13105}{72}\zeta_2- \frac{3049531}{31104} \right) C_F C_A^2  \nn \\
&+ \left(-432 \zeta_5-208 \zeta_2 \zeta_3+ \frac{8252}{3}\zeta_4- \frac{19424}{9} \zeta_3- \frac{16709}{27}\zeta_2+\frac{20329835}{15552} \right) C_F^2 C_A \nn \\
&+\left(416\zeta_5+224\zeta_2\zeta_3- \frac{6172}{3}\zeta_4+\frac{10942}{9}\zeta_3+\frac{11797}{18}\zeta_2- \frac{17471825}{15552} \right) C_F^3 \nn \\
&+\left(\frac{68}{3}\zeta_4-\frac{5803}{45}\zeta_3 + \frac{146971}{2700}\zeta_2-\frac{25234031}{1944000} \right) C_A C_F n_f
+\left(-\frac{136}{3}\zeta_4+\frac{8176}{45}\zeta_3-\frac{9767}{225}\zeta_2-\frac{4100189}{64800} \right) C_F^2 n_f \nn \\
&-\frac{105799}{162000} \, C_F n_f^2 \,, \nn \\
\gamma_{T,gq}^{(2)} &=
\left(\frac{196}{3}\zeta_4-\frac{2791}{90}\zeta_3-\frac{50593}{600}\zeta_2-\frac{17093053}{777600} \right) C_F C_A^2 + \left(\frac{511}{3}\zeta_4-\frac{3029}{9}\zeta_3+\frac{123773}{900}\zeta_2+\frac{63294389}{388800} \right) C_F^2 C_A \nn \\
&+ \left(-308\zeta_4+\frac{2533}{9}\zeta_3+\frac{3193}{54}\zeta_2-\frac{647639}{3888} \right) C_F^3 + \left(\frac{182}{9}\zeta_3-\frac{73}{27}\zeta_2+\frac{246767}{60750} \right)C_A C_F n_f \nn \\
&+ \left(-\frac{28}{9}\zeta_3+\frac{4}{9}\zeta_2-\frac{419593}{81000} \right) C_F^2 n_f\,, \nn \\
\gamma_{T,qg}^{(2)} &=
\left(-\frac{252}{5}\zeta_4+\frac{343}{45}\zeta_3+\frac{239959}{13500}\zeta_2-\frac{1795237}{1944000} \right) C_A^2 n_f
+\left(-\frac{42}{5}\zeta_4+\frac{6208}{75}\zeta_3+\frac{34127}{1350}\zeta_2-\frac{3607891}{38880} \right) C_A C_F n_f \nn \\
&+\left(\frac{448}{15}\zeta_4-\frac{26102}{225}\zeta_3-\frac{2042}{225}\zeta_2+\frac{9397651}{97200}\right) C_F^2 n_f
+\left(-\frac{28}{9}\zeta_3-\frac{554}{135}\zeta_2+\frac{1215691}{121500}\right) C_A n_f^2 \nn \\
&+ \left(\frac{2738}{675}\zeta_2-\frac{10657}{4050}\right) C_F n_f^2
 - \frac{172}{1125} \, n_f^3\,, \nn \\
\gamma_{T,gg}^{(2)} &=
\left(96\zeta_5+64\zeta_2 \zeta_3-\frac{2566}{15}\zeta_4-\frac{23702}{225}\zeta_3+\frac{66358}{1125}\zeta_2-\frac{5819653}{486000} \right)C_A^3 \nn \\
&+\left(104 \zeta_4+\frac{239}{9}\zeta_3-\frac{51269}{540}\zeta_2-\frac{12230737}{1944000} \right) C_A^2 n_f \nn \\
&+\left(\frac{282}{5}\zeta_3-\frac{16291}{675}\zeta_2-\frac{1700563}{108000} \right) C_A C_F n_f
+\left(-\frac{28}{9}\zeta_3+\frac{2411}{675}\zeta_2+\frac{219077}{194400} \right)C_F^2 n_f \nn \\
&+\left(-\frac{64}{9}\zeta_3+\frac{160}{27}\zeta_2-\frac{18269}{10125} \right) C_A n_f^2+\left(-\frac{196}{135}\zeta_2-\frac{2611}{162000} \right) C_F n_f^2\,.
\end{align}
Beyond LL, due to the appearance of $\ln y$ in the jet function on the right-hand side of the RG equation~\eqref{eq:jetRG}, one encounters the same moments of the splitting functions, but weighted by additional logarithms,
\begin{align}
\partial^n_N \gamma_{T,ij}^{(L)}\ =\  - \int\limits_0^1 dx \, x^2 \, \ln^n \!x \, P^{(L)}_{ij}(x) \,.
\end{align}
We have used this notation since these logarithmic moments correspond to Mellin space derivatives, evaluated at $N=3$, namely  
\begin{align}
\int\limits_0^1 dx \, x^{N-1} \ln^n \!x \, P^{(L)}_{ij}(x)
\ =\ \frac{\partial^n}{\partial N^n} 
\int\limits_0^1 dx \, x^{N-1} \, P^{(L)}_{ij}(x) \,.
\end{align}
We also use the shorthand $\dot\gamma \equiv \partial_N \gamma$
and $\ddot\gamma \equiv \partial^2_N \gamma$.
To NNLL, we require the first and second logarithmic moments of the LO splitting functions, and the first logarithmic moments of the NLO splitting functions. The logarithmic moments of the LO splitting functions are
\begin{align}
\dot{\gamma}_{T,qq}^{(0)}&= \left( 4\zeta_2 - \frac{385}{72} \right) C_F \,, \quad
\dot{\gamma}_{T,gq}^{(0)}= \frac{49}{72} \, C_F \,, \quad
\dot{\gamma}_{T,qg}^{(0)} = \frac{119}{900} \, n_f \,, \quad
\dot{\gamma}_{T,gg}^{(0)} = \left( 4\zeta_2 - \frac{4319}{900} \right) C_A \,,   \\
\ddot{\gamma}_{T,qq}^{(0)}&= \left( - 8\zeta_3 + \frac{3979}{432} \right) C_F \,, \quad
\ddot{\gamma}_{T,gq}^{(0)}= - \frac{331}{432} \, C_F \,, \quad
\ddot{\gamma}_{T,qg}^{(0)}= - \frac{2353}{27000} \, n_f  \,, \quad
\ddot{\gamma}_{T,gg}^{(0)}= \left( - 8\zeta_3 + \frac{230353}{27000} \right) C_A \,. \nn
\end{align}
The first logarithmic moments of the NLO splitting functions are
\begin{align}
\dot{\gamma}_{T,qq}^{(1)}&= \left(-56\zeta_4- \frac{158}{3}\zeta_3+\frac{385}{18}\zeta_2+\frac{152863}{1728}\right) C_F^2
+ \left(-12\zeta_4+\frac{41}{3}\zeta_3+\frac{307}{6}\zeta_2-\frac{35785}{432} \right) C_F C_A  \nn \\
&+ \left(\frac{16}{3}\zeta_3-\frac{40}{9}\zeta_2-\frac{101923}{108000} \right)  C_F n_f \,, \nn \\
\dot{\gamma}_{T,gq}^{(1)}&= \left(-\frac{49}{3}\zeta_3+\frac{59}{6}\zeta_2+\frac{956963}{108000} \right) C_F C_A
+ \left(14\zeta_3-\frac{275}{18}\zeta_2+\frac{8053}{1728} \right) C_F^2 \,, \nn \\
\dot{\gamma}_{T,qg}^{(1)} &= \left(\frac{42}{5}\zeta_3-\frac{92}{75}\zeta_2-\frac{1460321}{162000} \right) C_A n_f
+ \left(-\frac{28}{3}\zeta_3+\frac{178}{225}\zeta_2+\frac{46663}{4320} \right) C_F n_f  
+ \left(-\frac{28}{45}\zeta_2+\frac{18451}{20250} \right) n_f^2 \,, \nn \\
\dot{\gamma}_{T,gg}^{(1)} &= \left(-68\zeta_4-\frac{686}{15}\zeta_3+\frac{15338}{225}\zeta_2+\frac{3642257}{162000} \right) C_A^2 
+ \left(\frac{32}{3}\zeta_3-\frac{40}{9}\zeta_2-\frac{137323}{20250} \right) C_A n_f
- \frac{58247}{108000} \, C_F n_f .
\end{align}

We also record the hard function constants at $\mu=Q$ that are 
required for the $e^+e^-$ annihilation and Higgs decay processes,
extracted from refs.~\cite{Mitov:2006ic,Moch:2007tx,Almasy:2011eq}.
Again the $N=3$ moment is required at the first order
the hard coefficient appears, and integrals weighted with additional
powers of $\ln x$, again denoted by dots, appear at subsequent logarithmic
orders.  The Born level hard function does not require dots because it
is a delta function at $x=1$, and
$\int_0^1 dx \, x^2 \, \ln^nx \, \delta(1-x) = 0$ for $n>0$.
The coefficients required for $e^+e^-$ annihilation are defined as
\begin{align}
\int_0^1 dx \, x^2 \, H_{q,g}(x,\mu=Q)\ &=\ \sum\limits_{L=0}^\infty
\left( \frac{\alpha_s}{4\pi} \right)^{L} h_L^{q,g} \,, \nn\\
\int_0^1 dx \, x^2 \, \ln x \, H_{q,g}(x,\mu=Q)\ &=\ \sum\limits_{L=1}^\infty
\left( \frac{\alpha_s}{4\pi} \right)^{L} \dot{h}_L^{q,g} \,,
\end{align}
and so on.  The ones needed to NNLL are given by
\begin{align}
h_0^q &= \frac{1}{2} \,, \qquad
h_0^g = 0 \,, \qquad\qquad
h_1^q = \frac{131}{16} \, C_F \,, \qquad
h_1^g = - \frac{71}{48} \, C_F \,, \nn\\
h_2^q &= \left( 16 \zeta_4 - \frac{293}{3} \zeta_3 - \frac{83}{2} \zeta_2
              + \frac{2386397}{10368} \right) C_A C_F
+ \left( - 32 \zeta_4 + \frac{254}{3} \zeta_3 + \frac{1751}{72} \zeta_2
         - \frac{1105289}{20736} \right) C_F^2 \nn\\
& \hskip0.5cm\null
+ \left( 4 \zeta_3 + \frac{59}{60} \zeta_2 - \frac{8530817}{432000} \right)
   C_F n_f \,, \nn\\
h_2^g &= \left( - \frac{19}{3} \zeta_3 + \frac{47}{45} \zeta_2
   - \frac{29802739}{1296000} \right) C_A C_F
+ \left( \frac{31}{3} \zeta_3 + \frac{523}{72} \zeta_2
        - \frac{674045}{20736} \right) C_F^2 \,, \nn\\
\dot{h}_1^q &= \left( 10 \zeta_3 + \frac{61}{12}  \zeta_2
       - \frac{5303}{288} \right) C_F  \,, \qquad
\dot{h}_1^g = \left( - \frac{7}{12} \zeta_2 + \frac{31}{16} \right) C_F \,. 
\end{align}
We denote the coefficients required for the Higgs EEC with a capital $H$ instead
of a small $h$; they are given by
\begin{align}
H_0^q &= 0 \,, \qquad
H_0^g = \frac{1}{2} \,, \qquad\qquad
H_1^q = - \frac{163}{200} \, n_f \,, \qquad
H_1^g = \frac{5107}{300} \, C_A - \frac{79}{60} \, n_f \,, \nn\\
H_2^q &= \left( \frac{2743}{450} \zeta_2 - \frac{845983}{25920} \right) C_A n_f
+ \left( \frac{14}{15} \zeta_3 - \frac{73}{36} \zeta_2
       - \frac{575293}{51840} \right)  C_F n_f
+ \left( - \frac{28}{45} \zeta_2 + \frac{44396}{10125} \right) n_f^2  \,, \nn\\
H_2^g &= \left( - 16 \zeta_4 - \frac{469}{15} \zeta_3
        - \frac{12314}{225} \zeta_2 + \frac{19217009}{36000} \right) C_A^2
+ \left( - \frac{26}{3} \zeta_3 + \frac{137}{15} \zeta_2
         - \frac{33580213}{324000} \right) C_A n_f \nn\\
& \hskip0.5cm\null
+ \left( 12 \zeta_3 - \frac{49}{180} \zeta_2
       - \frac{20736797}{1296000} \right) C_F n_f
+ \left( - \frac{4}{9} \zeta_2 + \frac{30719}{8100} \right) n_f^2 \,, \nn\\
\dot{H}_1^q &= \left( - \frac{7}{30} \zeta_2 + \frac{4999}{9000} \right) n_f
\,, \qquad
\dot{H}_1^g = \left( 10 \zeta_3 + \frac{76}{15} \zeta_2
     - \frac{1905163}{108000} \right) C_A
+ \left( - \frac{1}{3} \zeta_2 + \frac{5269}{10800} \right) n_f \,.
\end{align}
Compared with refs.~\cite{Mitov:2006ic,Moch:2007tx,Almasy:2011eq}, we
require an overall factor of $1/2$ in three cases ($h^q$, $H^q$ and $H^g$),
and $1/4$ in the fourth case ($h^g$).
The factor of $E_i E_j/Q^2$ in the definition of the EEC gives rise
to a factor of $1/4$ because partons with Born kinematics have 
$E_i=E_j=Q/2$.  However, in most cases there is an additional factor of 2
because, for example, quarks and anti-quarks are summed in the EEC,
and are usually considered separately in fragmentation.
Also, for $e^+e^-$, because we integrate
over the incoming beam orientation, we use the sum of the transverse ($T$)
and longitudinal ($L$) hard functions.

\end{widetext}


\bibliography{EEC_forward.bib}{}

\begin{thebibliography}{91}%
\makeatletter
\providecommand \@ifxundefined [1]{%
 \@ifx{#1\undefined}
}%
\providecommand \@ifnum [1]{%
 \ifnum #1\expandafter \@firstoftwo
 \else \expandafter \@secondoftwo
 \fi
}%
\providecommand \@ifx [1]{%
 \ifx #1\expandafter \@firstoftwo
 \else \expandafter \@secondoftwo
 \fi
}%
\providecommand \natexlab [1]{#1}%
\providecommand \enquote  [1]{``#1''}%
\providecommand \bibnamefont  [1]{#1}%
\providecommand \bibfnamefont [1]{#1}%
\providecommand \citenamefont [1]{#1}%
\providecommand \href@noop [0]{\@secondoftwo}%
\providecommand \href [0]{\begingroup \@sanitize@url \@href}%
\providecommand \@href[1]{\@@startlink{#1}\@@href}%
\providecommand \@@href[1]{\endgroup#1\@@endlink}%
\providecommand \@sanitize@url [0]{\catcode `\\12\catcode `\$12\catcode
  `\&12\catcode `\#12\catcode `\^12\catcode `\_12\catcode `\%12\relax}%
\providecommand \@@startlink[1]{}%
\providecommand \@@endlink[0]{}%
\providecommand \url  [0]{\begingroup\@sanitize@url \@url }%
\providecommand \@url [1]{\endgroup\@href {#1}{\urlprefix }}%
\providecommand \urlprefix  [0]{URL }%
\providecommand \Eprint [0]{\href }%
\providecommand \doibase [0]{http://dx.doi.org/}%
\providecommand \selectlanguage [0]{\@gobble}%
\providecommand \bibinfo  [0]{\@secondoftwo}%
\providecommand \bibfield  [0]{\@secondoftwo}%
\providecommand \translation [1]{[#1]}%
\providecommand \BibitemOpen [0]{}%
\providecommand \bibitemStop [0]{}%
\providecommand \bibitemNoStop [0]{.\EOS\space}%
\providecommand \EOS [0]{\spacefactor3000\relax}%
\providecommand \BibitemShut  [1]{\csname bibitem#1\endcsname}%
\let\auto@bib@innerbib\@empty
\bibitem [{\citenamefont {Basham}\ \emph {et~al.}(1978)\citenamefont {Basham},
  \citenamefont {Brown}, \citenamefont {Ellis},\ and\ \citenamefont
  {Love}}]{Basham:1978bw}%
  \BibitemOpen
  \bibfield  {author} {\bibinfo {author} {\bibfnamefont {C.~L.}\ \bibnamefont
  {Basham}}, \bibinfo {author} {\bibfnamefont {L.~S.}\ \bibnamefont {Brown}},
  \bibinfo {author} {\bibfnamefont {S.~D.}\ \bibnamefont {Ellis}}, \ and\
  \bibinfo {author} {\bibfnamefont {S.~T.}\ \bibnamefont {Love}},\ }\href
  {\doibase 10.1103/PhysRevLett.41.1585} {\bibfield  {journal} {\bibinfo
  {journal} {Phys. Rev. Lett.}\ }\textbf {\bibinfo {volume} {41}},\ \bibinfo
  {pages} {1585} (\bibinfo {year} {1978})}\BibitemShut {NoStop}%
\bibitem [{\citenamefont {Basham}\ \emph {et~al.}(1979)\citenamefont {Basham},
  \citenamefont {Brown}, \citenamefont {Ellis},\ and\ \citenamefont
  {Love}}]{Basham:1978zq}%
  \BibitemOpen
  \bibfield  {author} {\bibinfo {author} {\bibfnamefont {C.~L.}\ \bibnamefont
  {Basham}}, \bibinfo {author} {\bibfnamefont {L.~S.}\ \bibnamefont {Brown}},
  \bibinfo {author} {\bibfnamefont {S.~D.}\ \bibnamefont {Ellis}}, \ and\
  \bibinfo {author} {\bibfnamefont {S.~T.}\ \bibnamefont {Love}},\ }\href
  {\doibase 10.1103/PhysRevD.19.2018} {\bibfield  {journal} {\bibinfo
  {journal} {Phys. Rev.}\ }\textbf {\bibinfo {volume} {D19}},\ \bibinfo {pages}
  {2018} (\bibinfo {year} {1979})}\BibitemShut {NoStop}%
\bibitem [{\citenamefont {Luo}\ \emph {et~al.}(2019)\citenamefont {Luo},
  \citenamefont {Shtabovenko}, \citenamefont {Yang},\ and\ \citenamefont
  {Zhu}}]{Luo:2019nig}%
  \BibitemOpen
  \bibfield  {author} {\bibinfo {author} {\bibfnamefont {M.-x.}\ \bibnamefont
  {Luo}}, \bibinfo {author} {\bibfnamefont {V.}~\bibnamefont {Shtabovenko}},
  \bibinfo {author} {\bibfnamefont {T.-Z.}\ \bibnamefont {Yang}}, \ and\
  \bibinfo {author} {\bibfnamefont {H.~X.}\ \bibnamefont {Zhu}},\ }\href@noop
  {} {\  (\bibinfo {year} {2019})},\ \Eprint {http://arxiv.org/abs/1903.07277}
  {arXiv:1903.07277 [hep-ph]} \BibitemShut {NoStop}%
\bibitem [{\citenamefont {Hofman}\ and\ \citenamefont
  {Maldacena}(2008)}]{Hofman:2008ar}%
  \BibitemOpen
  \bibfield  {author} {\bibinfo {author} {\bibfnamefont {D.~M.}\ \bibnamefont
  {Hofman}}\ and\ \bibinfo {author} {\bibfnamefont {J.}~\bibnamefont
  {Maldacena}},\ }\href {\doibase 10.1088/1126-6708/2008/05/012} {\bibfield
  {journal} {\bibinfo  {journal} {JHEP}\ }\textbf {\bibinfo {volume} {05}},\
  \bibinfo {pages} {012} (\bibinfo {year} {2008})},\ \Eprint
  {http://arxiv.org/abs/0803.1467} {arXiv:0803.1467 [hep-th]} \BibitemShut
  {NoStop}%
\bibitem [{\citenamefont {Belitsky}\ \emph
  {et~al.}(2014{\natexlab{a}})\citenamefont {Belitsky}, \citenamefont
  {Hohenegger}, \citenamefont {Korchemsky}, \citenamefont {Sokatchev},\ and\
  \citenamefont {Zhiboedov}}]{Belitsky:2013xxa}%
  \BibitemOpen
  \bibfield  {author} {\bibinfo {author} {\bibfnamefont {A.~V.}\ \bibnamefont
  {Belitsky}}, \bibinfo {author} {\bibfnamefont {S.}~\bibnamefont
  {Hohenegger}}, \bibinfo {author} {\bibfnamefont {G.~P.}\ \bibnamefont
  {Korchemsky}}, \bibinfo {author} {\bibfnamefont {E.}~\bibnamefont
  {Sokatchev}}, \ and\ \bibinfo {author} {\bibfnamefont {A.}~\bibnamefont
  {Zhiboedov}},\ }\href {\doibase 10.1016/j.nuclphysb.2014.04.020} {\bibfield
  {journal} {\bibinfo  {journal} {Nucl. Phys.}\ }\textbf {\bibinfo {volume}
  {B884}},\ \bibinfo {pages} {305} (\bibinfo {year} {2014}{\natexlab{a}})},\
  \Eprint {http://arxiv.org/abs/1309.0769} {arXiv:1309.0769 [hep-th]}
  \BibitemShut {NoStop}%
\bibitem [{\citenamefont {Belitsky}\ \emph
  {et~al.}(2014{\natexlab{b}})\citenamefont {Belitsky}, \citenamefont
  {Hohenegger}, \citenamefont {Korchemsky}, \citenamefont {Sokatchev},\ and\
  \citenamefont {Zhiboedov}}]{Belitsky:2013bja}%
  \BibitemOpen
  \bibfield  {author} {\bibinfo {author} {\bibfnamefont {A.~V.}\ \bibnamefont
  {Belitsky}}, \bibinfo {author} {\bibfnamefont {S.}~\bibnamefont
  {Hohenegger}}, \bibinfo {author} {\bibfnamefont {G.~P.}\ \bibnamefont
  {Korchemsky}}, \bibinfo {author} {\bibfnamefont {E.}~\bibnamefont
  {Sokatchev}}, \ and\ \bibinfo {author} {\bibfnamefont {A.}~\bibnamefont
  {Zhiboedov}},\ }\href {\doibase 10.1016/j.nuclphysb.2014.04.019} {\bibfield
  {journal} {\bibinfo  {journal} {Nucl. Phys.}\ }\textbf {\bibinfo {volume}
  {B884}},\ \bibinfo {pages} {206} (\bibinfo {year} {2014}{\natexlab{b}})},\
  \Eprint {http://arxiv.org/abs/1309.1424} {arXiv:1309.1424 [hep-th]}
  \BibitemShut {NoStop}%
\bibitem [{\citenamefont {Belitsky}\ \emph
  {et~al.}(2014{\natexlab{c}})\citenamefont {Belitsky}, \citenamefont
  {Hohenegger}, \citenamefont {Korchemsky}, \citenamefont {Sokatchev},\ and\
  \citenamefont {Zhiboedov}}]{Belitsky:2013ofa}%
  \BibitemOpen
  \bibfield  {author} {\bibinfo {author} {\bibfnamefont {A.~V.}\ \bibnamefont
  {Belitsky}}, \bibinfo {author} {\bibfnamefont {S.}~\bibnamefont
  {Hohenegger}}, \bibinfo {author} {\bibfnamefont {G.~P.}\ \bibnamefont
  {Korchemsky}}, \bibinfo {author} {\bibfnamefont {E.}~\bibnamefont
  {Sokatchev}}, \ and\ \bibinfo {author} {\bibfnamefont {A.}~\bibnamefont
  {Zhiboedov}},\ }\href {\doibase 10.1103/PhysRevLett.112.071601} {\bibfield
  {journal} {\bibinfo  {journal} {Phys. Rev. Lett.}\ }\textbf {\bibinfo
  {volume} {112}},\ \bibinfo {pages} {071601} (\bibinfo {year}
  {2014}{\natexlab{c}})},\ \Eprint {http://arxiv.org/abs/1311.6800}
  {arXiv:1311.6800 [hep-th]} \BibitemShut {NoStop}%
\bibitem [{\citenamefont {Dixon}\ \emph {et~al.}(2018)\citenamefont {Dixon},
  \citenamefont {Luo}, \citenamefont {Shtabovenko}, \citenamefont {Yang},\ and\
  \citenamefont {Zhu}}]{Dixon:2018qgp}%
  \BibitemOpen
  \bibfield  {author} {\bibinfo {author} {\bibfnamefont {L.~J.}\ \bibnamefont
  {Dixon}}, \bibinfo {author} {\bibfnamefont {M.-X.}\ \bibnamefont {Luo}},
  \bibinfo {author} {\bibfnamefont {V.}~\bibnamefont {Shtabovenko}}, \bibinfo
  {author} {\bibfnamefont {T.-Z.}\ \bibnamefont {Yang}}, \ and\ \bibinfo
  {author} {\bibfnamefont {H.~X.}\ \bibnamefont {Zhu}},\ }\href {\doibase
  10.1103/PhysRevLett.120.102001} {\bibfield  {journal} {\bibinfo  {journal}
  {Phys. Rev. Lett.}\ }\textbf {\bibinfo {volume} {120}},\ \bibinfo {pages}
  {102001} (\bibinfo {year} {2018})},\ \Eprint
  {http://arxiv.org/abs/1801.03219} {arXiv:1801.03219 [hep-ph]} \BibitemShut
  {NoStop}%
\bibitem [{\citenamefont {Henn}\ \emph {et~al.}(2019)\citenamefont {Henn},
  \citenamefont {Sokatchev}, \citenamefont {Yan},\ and\ \citenamefont
  {Zhiboedov}}]{Henn:2019gkr}%
  \BibitemOpen
  \bibfield  {author} {\bibinfo {author} {\bibfnamefont {J.~M.}\ \bibnamefont
  {Henn}}, \bibinfo {author} {\bibfnamefont {E.}~\bibnamefont {Sokatchev}},
  \bibinfo {author} {\bibfnamefont {K.}~\bibnamefont {Yan}}, \ and\ \bibinfo
  {author} {\bibfnamefont {A.}~\bibnamefont {Zhiboedov}},\ }\href@noop {} {\
  (\bibinfo {year} {2019})},\ \Eprint {http://arxiv.org/abs/1903.05314}
  {arXiv:1903.05314 [hep-th]} \BibitemShut {NoStop}%
\bibitem [{\citenamefont {Moult}\ and\ \citenamefont
  {Zhu}(2018)}]{Moult:2018jzp}%
  \BibitemOpen
  \bibfield  {author} {\bibinfo {author} {\bibfnamefont {I.}~\bibnamefont
  {Moult}}\ and\ \bibinfo {author} {\bibfnamefont {H.~X.}\ \bibnamefont
  {Zhu}},\ }\href {\doibase 10.1007/JHEP08(2018)160} {\bibfield  {journal}
  {\bibinfo  {journal} {JHEP}\ }\textbf {\bibinfo {volume} {08}},\ \bibinfo
  {pages} {160} (\bibinfo {year} {2018})},\ \Eprint
  {http://arxiv.org/abs/1801.02627} {arXiv:1801.02627 [hep-ph]} \BibitemShut
  {NoStop}%
\bibitem [{\citenamefont {Gao}\ \emph {et~al.}(2019)\citenamefont {Gao},
  \citenamefont {Li}, \citenamefont {Moult},\ and\ \citenamefont
  {Zhu}}]{Gao:2019ojf}%
  \BibitemOpen
  \bibfield  {author} {\bibinfo {author} {\bibfnamefont {A.}~\bibnamefont
  {Gao}}, \bibinfo {author} {\bibfnamefont {H.~T.}\ \bibnamefont {Li}},
  \bibinfo {author} {\bibfnamefont {I.}~\bibnamefont {Moult}}, \ and\ \bibinfo
  {author} {\bibfnamefont {H.~X.}\ \bibnamefont {Zhu}},\ }\href@noop {} {\
  (\bibinfo {year} {2019})},\ \Eprint {http://arxiv.org/abs/1901.04497}
  {arXiv:1901.04497 [hep-ph]} \BibitemShut {NoStop}%
\bibitem [{\citenamefont {Del~Duca}\ \emph
  {et~al.}(2016{\natexlab{a}})\citenamefont {Del~Duca}, \citenamefont {Duhr},
  \citenamefont {Kardos}, \citenamefont {Somogyi}, \citenamefont {Sz{\H o}r},
  \citenamefont {Tr{\'o}cs{\'a}nyi},\ and\ \citenamefont
  {Tulip{\'a}nt}}]{DelDuca:2016ily}%
  \BibitemOpen
  \bibfield  {author} {\bibinfo {author} {\bibfnamefont {V.}~\bibnamefont
  {Del~Duca}}, \bibinfo {author} {\bibfnamefont {C.}~\bibnamefont {Duhr}},
  \bibinfo {author} {\bibfnamefont {A.}~\bibnamefont {Kardos}}, \bibinfo
  {author} {\bibfnamefont {G.}~\bibnamefont {Somogyi}}, \bibinfo {author}
  {\bibfnamefont {Z.}~\bibnamefont {Sz{\H o}r}}, \bibinfo {author}
  {\bibfnamefont {Z.}~\bibnamefont {Tr{\'o}cs{\'a}nyi}}, \ and\ \bibinfo
  {author} {\bibfnamefont {Z.}~\bibnamefont {Tulip{\'a}nt}},\ }\href {\doibase
  10.1103/PhysRevD.94.074019} {\bibfield  {journal} {\bibinfo  {journal} {Phys.
  Rev.}\ }\textbf {\bibinfo {volume} {D94}},\ \bibinfo {pages} {074019}
  (\bibinfo {year} {2016}{\natexlab{a}})},\ \Eprint
  {http://arxiv.org/abs/1606.03453} {arXiv:1606.03453 [hep-ph]} \BibitemShut
  {NoStop}%
\bibitem [{\citenamefont {{Tulip{\'a}nt, Z. and Kardos, A. and Somogyi,
  G.}}(2017)}]{Tulipant:2017ybb}%
  \BibitemOpen
  \bibfield  {author} {\bibinfo {author} {\bibnamefont {{Tulip{\'a}nt, Z. and
  Kardos, A. and Somogyi, G.}}},\ }\href {\doibase
  10.1140/epjc/s10052-017-5320-9} {\bibfield  {journal} {\bibinfo  {journal}
  {Eur. Phys. J.}\ }\textbf {\bibinfo {volume} {C77}},\ \bibinfo {pages} {749}
  (\bibinfo {year} {2017})},\ \Eprint {http://arxiv.org/abs/1708.04093}
  {arXiv:1708.04093 [hep-ph]} \BibitemShut {NoStop}%
\bibitem [{\citenamefont {de~Florian}\ and\ \citenamefont
  {Grazzini}(2005)}]{deFlorian:2004mp}%
  \BibitemOpen
  \bibfield  {author} {\bibinfo {author} {\bibfnamefont {D.}~\bibnamefont
  {de~Florian}}\ and\ \bibinfo {author} {\bibfnamefont {M.}~\bibnamefont
  {Grazzini}},\ }\href {\doibase 10.1016/j.nuclphysb.2004.10.051} {\bibfield
  {journal} {\bibinfo  {journal} {Nucl. Phys.}\ }\textbf {\bibinfo {volume}
  {B704}},\ \bibinfo {pages} {387} (\bibinfo {year} {2005})},\ \Eprint
  {http://arxiv.org/abs/hep-ph/0407241} {arXiv:hep-ph/0407241 [hep-ph]}
  \BibitemShut {NoStop}%
\bibitem [{\citenamefont {Kardos}\ \emph {et~al.}(2018)\citenamefont {Kardos},
  \citenamefont {Kluth}, \citenamefont {Somogyi}, \citenamefont
  {Tulip{\'a}nt},\ and\ \citenamefont {Verbytskyi}}]{Kardos:2018kqj}%
  \BibitemOpen
  \bibfield  {author} {\bibinfo {author} {\bibfnamefont {A.}~\bibnamefont
  {Kardos}}, \bibinfo {author} {\bibfnamefont {S.}~\bibnamefont {Kluth}},
  \bibinfo {author} {\bibfnamefont {G.}~\bibnamefont {Somogyi}}, \bibinfo
  {author} {\bibfnamefont {Z.}~\bibnamefont {Tulip{\'a}nt}}, \ and\ \bibinfo
  {author} {\bibfnamefont {A.}~\bibnamefont {Verbytskyi}},\ }\href@noop {} {\
  (\bibinfo {year} {2018})},\ \Eprint {http://arxiv.org/abs/1804.09146}
  {arXiv:1804.09146 [hep-ph]} \BibitemShut {NoStop}%
\bibitem [{\citenamefont {Simmons-Duffin}\ \emph {et~al.}()\citenamefont
  {Simmons-Duffin}, \citenamefont {Kravchuk},\ and\ \citenamefont
  {Zhiboedov}}]{DavidPetrSashaTalks}%
  \BibitemOpen
  \bibfield  {author} {\bibinfo {author} {\bibfnamefont {D.}~\bibnamefont
  {Simmons-Duffin}}, \bibinfo {author} {\bibfnamefont {P.}~\bibnamefont
  {Kravchuk}}, \ and\ \bibinfo {author} {\bibfnamefont {A.}~\bibnamefont
  {Zhiboedov}},\ }\href
  {http://online.kitp.ucsb.edu/online/polchinski_c18/simmonsduffin/} {\enquote
  {\bibinfo {title} {Seminars (2018 and 2019)},}\ }\bibinfo {howpublished}
  {\url{http://online.kitp.ucsb.edu/online/polchinski_c18/simmonsduffin/}}\BibitemShut
  {NoStop}%
\bibitem [{DSD()}]{DSDtocome}%
  \BibitemOpen
  \href@noop {} {}\bibinfo {howpublished} {M.~Kologlu, P.~Kravchuk,
  D.~Simmons-Duffin, and A.~Zhiboedov, {\it The light-ray OPE and conformal
  colliders}}\BibitemShut {NoStop}%
\bibitem [{\citenamefont {Kravchuk}\ and\ \citenamefont
  {Simmons-Duffin}(2018)}]{Kravchuk:2018htv}%
  \BibitemOpen
  \bibfield  {author} {\bibinfo {author} {\bibfnamefont {P.}~\bibnamefont
  {Kravchuk}}\ and\ \bibinfo {author} {\bibfnamefont {D.}~\bibnamefont
  {Simmons-Duffin}},\ }\href {\doibase 10.1007/JHEP11(2018)102} {\bibfield
  {journal} {\bibinfo  {journal} {JHEP}\ }\textbf {\bibinfo {volume} {11}},\
  \bibinfo {pages} {102} (\bibinfo {year} {2018})},\ \Eprint
  {http://arxiv.org/abs/1805.00098} {arXiv:1805.00098 [hep-th]} \BibitemShut
  {NoStop}%
\bibitem [{\citenamefont {Kologlu}\ \emph {et~al.}(2019)\citenamefont
  {Kologlu}, \citenamefont {Kravchuk}, \citenamefont {Simmons-Duffin},\ and\
  \citenamefont {Zhiboedov}}]{Kologlu:2019bco}%
  \BibitemOpen
  \bibfield  {author} {\bibinfo {author} {\bibfnamefont {M.}~\bibnamefont
  {Kologlu}}, \bibinfo {author} {\bibfnamefont {P.}~\bibnamefont {Kravchuk}},
  \bibinfo {author} {\bibfnamefont {D.}~\bibnamefont {Simmons-Duffin}}, \ and\
  \bibinfo {author} {\bibfnamefont {A.}~\bibnamefont {Zhiboedov}},\ }\href@noop
  {} {\  (\bibinfo {year} {2019})},\ \Eprint {http://arxiv.org/abs/1904.05905}
  {arXiv:1904.05905 [hep-th]} \BibitemShut {NoStop}%
\bibitem [{\citenamefont {Konishi}\ \emph
  {et~al.}(1979{\natexlab{a}})\citenamefont {Konishi}, \citenamefont {Ukawa},\
  and\ \citenamefont {Veneziano}}]{Konishi:1979cb}%
  \BibitemOpen
  \bibfield  {author} {\bibinfo {author} {\bibfnamefont {K.}~\bibnamefont
  {Konishi}}, \bibinfo {author} {\bibfnamefont {A.}~\bibnamefont {Ukawa}}, \
  and\ \bibinfo {author} {\bibfnamefont {G.}~\bibnamefont {Veneziano}},\ }\href
  {\doibase 10.1016/0550-3213(79)90053-1} {\bibfield  {journal} {\bibinfo
  {journal} {Nucl. Phys.}\ }\textbf {\bibinfo {volume} {B157}},\ \bibinfo
  {pages} {45} (\bibinfo {year} {1979}{\natexlab{a}})}\BibitemShut {NoStop}%
\bibitem [{Kor()}]{Korchemskytocome}%
  \BibitemOpen
  \href@noop {} {}\bibinfo {howpublished} {G.~P.~Korchemsky, {\it Energy
  correlations in the end-point region}, preprint IPhT--T19/041}\BibitemShut
  {NoStop}%
\bibitem [{\citenamefont {Konishi}\ \emph {et~al.}(1978)\citenamefont
  {Konishi}, \citenamefont {Ukawa},\ and\ \citenamefont
  {Veneziano}}]{Konishi:1978yx}%
  \BibitemOpen
  \bibfield  {author} {\bibinfo {author} {\bibfnamefont {K.}~\bibnamefont
  {Konishi}}, \bibinfo {author} {\bibfnamefont {A.}~\bibnamefont {Ukawa}}, \
  and\ \bibinfo {author} {\bibfnamefont {G.}~\bibnamefont {Veneziano}},\ }\href
  {\doibase 10.1016/0370-2693(78)90015-1} {\bibfield  {journal} {\bibinfo
  {journal} {Phys. Lett.}\ }\textbf {\bibinfo {volume} {78B}},\ \bibinfo
  {pages} {243} (\bibinfo {year} {1978})}\BibitemShut {NoStop}%
\bibitem [{\citenamefont {Konishi}\ \emph
  {et~al.}(1979{\natexlab{b}})\citenamefont {Konishi}, \citenamefont {Ukawa},\
  and\ \citenamefont {Veneziano}}]{Konishi:1978ax}%
  \BibitemOpen
  \bibfield  {author} {\bibinfo {author} {\bibfnamefont {K.}~\bibnamefont
  {Konishi}}, \bibinfo {author} {\bibfnamefont {A.}~\bibnamefont {Ukawa}}, \
  and\ \bibinfo {author} {\bibfnamefont {G.}~\bibnamefont {Veneziano}},\ }\href
  {\doibase 10.1016/0370-2693(79)90212-0} {\bibfield  {journal} {\bibinfo
  {journal} {Phys. Lett.}\ }\textbf {\bibinfo {volume} {80B}},\ \bibinfo
  {pages} {259} (\bibinfo {year} {1979}{\natexlab{b}})}\BibitemShut {NoStop}%
\bibitem [{\citenamefont {Kalinowski}\ \emph {et~al.}(1981)\citenamefont
  {Kalinowski}, \citenamefont {Konishi}, \citenamefont {Scharbach},\ and\
  \citenamefont {Taylor}}]{Kalinowski:1980wea}%
  \BibitemOpen
  \bibfield  {author} {\bibinfo {author} {\bibfnamefont {J.}~\bibnamefont
  {Kalinowski}}, \bibinfo {author} {\bibfnamefont {K.}~\bibnamefont {Konishi}},
  \bibinfo {author} {\bibfnamefont {P.~N.}\ \bibnamefont {Scharbach}}, \ and\
  \bibinfo {author} {\bibfnamefont {T.~R.}\ \bibnamefont {Taylor}},\ }\href
  {\doibase 10.1016/0550-3213(81)90352-7} {\bibfield  {journal} {\bibinfo
  {journal} {Nucl. Phys.}\ }\textbf {\bibinfo {volume} {B181}},\ \bibinfo
  {pages} {253} (\bibinfo {year} {1981})}\BibitemShut {NoStop}%
\bibitem [{\citenamefont {Richards}\ \emph {et~al.}(1982)\citenamefont
  {Richards}, \citenamefont {Stirling},\ and\ \citenamefont
  {Ellis}}]{Richards:1982te}%
  \BibitemOpen
  \bibfield  {author} {\bibinfo {author} {\bibfnamefont {D.~G.}\ \bibnamefont
  {Richards}}, \bibinfo {author} {\bibfnamefont {W.~J.}\ \bibnamefont
  {Stirling}}, \ and\ \bibinfo {author} {\bibfnamefont {S.~D.}\ \bibnamefont
  {Ellis}},\ }\href {\doibase 10.1016/0370-2693(82)90275-1} {\bibfield
  {journal} {\bibinfo  {journal} {Phys. Lett.}\ }\textbf {\bibinfo {volume}
  {119B}},\ \bibinfo {pages} {193} (\bibinfo {year} {1982})}\BibitemShut
  {NoStop}%
\bibitem [{\citenamefont {Rijken}\ and\ \citenamefont {van
  Neerven}(1997)}]{Rijken:1996ns}%
  \BibitemOpen
  \bibfield  {author} {\bibinfo {author} {\bibfnamefont {P.~J.}\ \bibnamefont
  {Rijken}}\ and\ \bibinfo {author} {\bibfnamefont {W.~L.}\ \bibnamefont {van
  Neerven}},\ }\href {\doibase 10.1016/S0550-3213(96)00669-4} {\bibfield
  {journal} {\bibinfo  {journal} {Nucl. Phys.}\ }\textbf {\bibinfo {volume}
  {B487}},\ \bibinfo {pages} {233} (\bibinfo {year} {1997})},\ \Eprint
  {http://arxiv.org/abs/hep-ph/9609377} {arXiv:hep-ph/9609377 [hep-ph]}
  \BibitemShut {NoStop}%
\bibitem [{\citenamefont {Mitov}\ and\ \citenamefont
  {Moch}(2006)}]{Mitov:2006wy}%
  \BibitemOpen
  \bibfield  {author} {\bibinfo {author} {\bibfnamefont {A.}~\bibnamefont
  {Mitov}}\ and\ \bibinfo {author} {\bibfnamefont {S.}~\bibnamefont {Moch}},\
  }\href {\doibase 10.1016/j.nuclphysb.2006.05.018} {\bibfield  {journal}
  {\bibinfo  {journal} {Nucl. Phys.}\ }\textbf {\bibinfo {volume} {B751}},\
  \bibinfo {pages} {18} (\bibinfo {year} {2006})},\ \Eprint
  {http://arxiv.org/abs/hep-ph/0604160} {arXiv:hep-ph/0604160 [hep-ph]}
  \BibitemShut {NoStop}%
\bibitem [{\citenamefont {Mitov}\ \emph {et~al.}(2006)\citenamefont {Mitov},
  \citenamefont {Moch},\ and\ \citenamefont {Vogt}}]{Mitov:2006ic}%
  \BibitemOpen
  \bibfield  {author} {\bibinfo {author} {\bibfnamefont {A.}~\bibnamefont
  {Mitov}}, \bibinfo {author} {\bibfnamefont {S.}~\bibnamefont {Moch}}, \ and\
  \bibinfo {author} {\bibfnamefont {A.}~\bibnamefont {Vogt}},\ }\href {\doibase
  10.1016/j.physletb.2006.05.005} {\bibfield  {journal} {\bibinfo  {journal}
  {Phys. Lett.}\ }\textbf {\bibinfo {volume} {B638}},\ \bibinfo {pages} {61}
  (\bibinfo {year} {2006})},\ \Eprint {http://arxiv.org/abs/hep-ph/0604053}
  {arXiv:hep-ph/0604053 [hep-ph]} \BibitemShut {NoStop}%
\bibitem [{\citenamefont {Moch}\ and\ \citenamefont
  {Vogt}(2008)}]{Moch:2007tx}%
  \BibitemOpen
  \bibfield  {author} {\bibinfo {author} {\bibfnamefont {S.}~\bibnamefont
  {Moch}}\ and\ \bibinfo {author} {\bibfnamefont {A.}~\bibnamefont {Vogt}},\
  }\href {\doibase 10.1016/j.physletb.2007.10.069} {\bibfield  {journal}
  {\bibinfo  {journal} {Phys. Lett.}\ }\textbf {\bibinfo {volume} {B659}},\
  \bibinfo {pages} {290} (\bibinfo {year} {2008})},\ \Eprint
  {http://arxiv.org/abs/0709.3899} {arXiv:0709.3899 [hep-ph]} \BibitemShut
  {NoStop}%
\bibitem [{\citenamefont {Almasy}\ \emph {et~al.}(2012)\citenamefont {Almasy},
  \citenamefont {Moch},\ and\ \citenamefont {Vogt}}]{Almasy:2011eq}%
  \BibitemOpen
  \bibfield  {author} {\bibinfo {author} {\bibfnamefont {A.~A.}\ \bibnamefont
  {Almasy}}, \bibinfo {author} {\bibfnamefont {S.}~\bibnamefont {Moch}}, \ and\
  \bibinfo {author} {\bibfnamefont {A.}~\bibnamefont {Vogt}},\ }\href {\doibase
  10.1016/j.nuclphysb.2011.08.028} {\bibfield  {journal} {\bibinfo  {journal}
  {Nucl. Phys.}\ }\textbf {\bibinfo {volume} {B854}},\ \bibinfo {pages} {133}
  (\bibinfo {year} {2012})},\ \Eprint {http://arxiv.org/abs/1107.2263}
  {arXiv:1107.2263 [hep-ph]} \BibitemShut {NoStop}%
\bibitem [{\citenamefont {Drell}\ \emph {et~al.}(1969)\citenamefont {Drell},
  \citenamefont {Levy},\ and\ \citenamefont {Yan}}]{Drell:1969jm}%
  \BibitemOpen
  \bibfield  {author} {\bibinfo {author} {\bibfnamefont {S.~D.}\ \bibnamefont
  {Drell}}, \bibinfo {author} {\bibfnamefont {D.~J.}\ \bibnamefont {Levy}}, \
  and\ \bibinfo {author} {\bibfnamefont {T.-M.}\ \bibnamefont {Yan}},\ }\href
  {\doibase 10.1103/PhysRev.187.2159} {\bibfield  {journal} {\bibinfo
  {journal} {Phys. Rev.}\ }\textbf {\bibinfo {volume} {187}},\ \bibinfo {pages}
  {2159} (\bibinfo {year} {1969})}\BibitemShut {NoStop}%
\bibitem [{\citenamefont {Gribov}\ and\ \citenamefont
  {Lipatov}(1972)}]{Gribov:1972ri}%
  \BibitemOpen
  \bibfield  {author} {\bibinfo {author} {\bibfnamefont {V.~N.}\ \bibnamefont
  {Gribov}}\ and\ \bibinfo {author} {\bibfnamefont {L.~N.}\ \bibnamefont
  {Lipatov}},\ }\href@noop {} {\bibfield  {journal} {\bibinfo  {journal} {Sov.
  J. Nucl. Phys.}\ }\textbf {\bibinfo {volume} {15}},\ \bibinfo {pages} {438}
  (\bibinfo {year} {1972})},\ \bibinfo {note} {[Yad.
  Fiz.15,781(1972)]}\BibitemShut {NoStop}%
\bibitem [{\citenamefont {Mueller}(1983)}]{Mueller:1983js}%
  \BibitemOpen
  \bibfield  {author} {\bibinfo {author} {\bibfnamefont {A.~H.}\ \bibnamefont
  {Mueller}},\ }\href {\doibase 10.1016/0550-3213(83)90329-2} {\bibfield
  {journal} {\bibinfo  {journal} {Nucl. Phys.}\ }\textbf {\bibinfo {volume}
  {B228}},\ \bibinfo {pages} {351} (\bibinfo {year} {1983})}\BibitemShut
  {NoStop}%
\bibitem [{\citenamefont {Blumlein}\ \emph {et~al.}(2000)\citenamefont
  {Blumlein}, \citenamefont {Ravindran},\ and\ \citenamefont {van
  Neerven}}]{Blumlein:2000wh}%
  \BibitemOpen
  \bibfield  {author} {\bibinfo {author} {\bibfnamefont {J.}~\bibnamefont
  {Blumlein}}, \bibinfo {author} {\bibfnamefont {V.}~\bibnamefont {Ravindran}},
  \ and\ \bibinfo {author} {\bibfnamefont {W.~L.}\ \bibnamefont {van
  Neerven}},\ }\href {\doibase 10.1016/S0550-3213(00)00422-3} {\bibfield
  {journal} {\bibinfo  {journal} {Nucl. Phys.}\ }\textbf {\bibinfo {volume}
  {B586}},\ \bibinfo {pages} {349} (\bibinfo {year} {2000})},\ \Eprint
  {http://arxiv.org/abs/hep-ph/0004172} {arXiv:hep-ph/0004172 [hep-ph]}
  \BibitemShut {NoStop}%
\bibitem [{\citenamefont {Dokshitzer}\ \emph {et~al.}(2006)\citenamefont
  {Dokshitzer}, \citenamefont {Marchesini},\ and\ \citenamefont
  {Salam}}]{Dokshitzer:2005bf}%
  \BibitemOpen
  \bibfield  {author} {\bibinfo {author} {\bibfnamefont {{\relax Yu}.~L.}\
  \bibnamefont {Dokshitzer}}, \bibinfo {author} {\bibfnamefont
  {G.}~\bibnamefont {Marchesini}}, \ and\ \bibinfo {author} {\bibfnamefont
  {G.~P.}\ \bibnamefont {Salam}},\ }\href {\doibase
  10.1016/j.physletb.2006.02.023} {\bibfield  {journal} {\bibinfo  {journal}
  {Phys. Lett.}\ }\textbf {\bibinfo {volume} {B634}},\ \bibinfo {pages} {504}
  (\bibinfo {year} {2006})},\ \Eprint {http://arxiv.org/abs/hep-ph/0511302}
  {arXiv:hep-ph/0511302 [hep-ph]} \BibitemShut {NoStop}%
\bibitem [{\citenamefont {Marchesini}(2006)}]{Marchesini:2006ax}%
  \BibitemOpen
  \bibfield  {author} {\bibinfo {author} {\bibfnamefont {G.}~\bibnamefont
  {Marchesini}},\ }in\ \href@noop {} {\emph {\bibinfo {booktitle}
  {{Proceedings, 41st Rencontres de Moriond, 2006 QCD and High Energy Hadronic
  Interactions: La Thuile, Val d'Aoste, Italy, Mar 18-25, 2006}}}}\ (\bibinfo
  {year} {2006})\ pp.\ \bibinfo {pages} {137--142},\ \Eprint
  {http://arxiv.org/abs/hep-ph/0605262} {arXiv:hep-ph/0605262 [hep-ph]}
  \BibitemShut {NoStop}%
\bibitem [{\citenamefont {Basso}\ and\ \citenamefont
  {Korchemsky}(2007)}]{Basso:2006nk}%
  \BibitemOpen
  \bibfield  {author} {\bibinfo {author} {\bibfnamefont {B.}~\bibnamefont
  {Basso}}\ and\ \bibinfo {author} {\bibfnamefont {G.~P.}\ \bibnamefont
  {Korchemsky}},\ }\href {\doibase 10.1016/j.nuclphysb.2007.03.044} {\bibfield
  {journal} {\bibinfo  {journal} {Nucl. Phys.}\ }\textbf {\bibinfo {volume}
  {B775}},\ \bibinfo {pages} {1} (\bibinfo {year} {2007})},\ \Eprint
  {http://arxiv.org/abs/hep-th/0612247} {arXiv:hep-th/0612247 [hep-th]}
  \BibitemShut {NoStop}%
\bibitem [{\citenamefont {Dokshitzer}\ and\ \citenamefont
  {Marchesini}(2007)}]{Dokshitzer:2006nm}%
  \BibitemOpen
  \bibfield  {author} {\bibinfo {author} {\bibfnamefont {{\relax Yu}.~L.}\
  \bibnamefont {Dokshitzer}}\ and\ \bibinfo {author} {\bibfnamefont
  {G.}~\bibnamefont {Marchesini}},\ }\href {\doibase
  10.1016/j.physletb.2007.01.016} {\bibfield  {journal} {\bibinfo  {journal}
  {Phys. Lett.}\ }\textbf {\bibinfo {volume} {B646}},\ \bibinfo {pages} {189}
  (\bibinfo {year} {2007})},\ \Eprint {http://arxiv.org/abs/hep-th/0612248}
  {arXiv:hep-th/0612248 [hep-th]} \BibitemShut {NoStop}%
\bibitem [{\citenamefont {Jaffe}\ and\ \citenamefont
  {Ji}(1992)}]{Jaffe:1991ra}%
  \BibitemOpen
  \bibfield  {author} {\bibinfo {author} {\bibfnamefont {R.~L.}\ \bibnamefont
  {Jaffe}}\ and\ \bibinfo {author} {\bibfnamefont {X.-D.}\ \bibnamefont {Ji}},\
  }\href {\doibase 10.1016/0550-3213(92)90110-W} {\bibfield  {journal}
  {\bibinfo  {journal} {Nucl. Phys.}\ }\textbf {\bibinfo {volume} {B375}},\
  \bibinfo {pages} {527} (\bibinfo {year} {1992})}\BibitemShut {NoStop}%
\bibitem [{\citenamefont {Bauer}\ \emph {et~al.}(2000)\citenamefont {Bauer},
  \citenamefont {Fleming},\ and\ \citenamefont {Luke}}]{Bauer:2000ew}%
  \BibitemOpen
  \bibfield  {author} {\bibinfo {author} {\bibfnamefont {C.~W.}\ \bibnamefont
  {Bauer}}, \bibinfo {author} {\bibfnamefont {S.}~\bibnamefont {Fleming}}, \
  and\ \bibinfo {author} {\bibfnamefont {M.~E.}\ \bibnamefont {Luke}},\ }\href
  {\doibase 10.1103/PhysRevD.63.014006} {\bibfield  {journal} {\bibinfo
  {journal} {Phys. Rev. D}\ }\textbf {\bibinfo {volume} {63}},\ \bibinfo
  {pages} {014006} (\bibinfo {year} {2000})},\ \Eprint
  {http://arxiv.org/abs/hep-ph/0005275} {hep-ph/0005275} \BibitemShut {NoStop}%
\bibitem [{\citenamefont {Bauer}\ \emph {et~al.}(2001)\citenamefont {Bauer},
  \citenamefont {Fleming}, \citenamefont {Pirjol},\ and\ \citenamefont
  {Stewart}}]{Bauer:2000yr}%
  \BibitemOpen
  \bibfield  {author} {\bibinfo {author} {\bibfnamefont {C.~W.}\ \bibnamefont
  {Bauer}}, \bibinfo {author} {\bibfnamefont {S.}~\bibnamefont {Fleming}},
  \bibinfo {author} {\bibfnamefont {D.}~\bibnamefont {Pirjol}}, \ and\ \bibinfo
  {author} {\bibfnamefont {I.~W.}\ \bibnamefont {Stewart}},\ }\href {\doibase
  10.1103/PhysRevD.63.114020} {\bibfield  {journal} {\bibinfo  {journal} {Phys.
  Rev. D}\ }\textbf {\bibinfo {volume} {63}},\ \bibinfo {pages} {114020}
  (\bibinfo {year} {2001})},\ \Eprint {http://arxiv.org/abs/hep-ph/0011336}
  {hep-ph/0011336} \BibitemShut {NoStop}%
\bibitem [{\citenamefont {Bauer}\ and\ \citenamefont
  {Stewart}(2001)}]{Bauer:2001ct}%
  \BibitemOpen
  \bibfield  {author} {\bibinfo {author} {\bibfnamefont {C.~W.}\ \bibnamefont
  {Bauer}}\ and\ \bibinfo {author} {\bibfnamefont {I.~W.}\ \bibnamefont
  {Stewart}},\ }\href {\doibase 10.1016/S0370-2693(01)00902-9} {\bibfield
  {journal} {\bibinfo  {journal} {Phys. Lett.}\ }\textbf {\bibinfo {volume}
  {B516}},\ \bibinfo {pages} {134} (\bibinfo {year} {2001})},\ \Eprint
  {http://arxiv.org/abs/hep-ph/0107001} {hep-ph/0107001} \BibitemShut {NoStop}%
\bibitem [{\citenamefont {Bauer}\ \emph {et~al.}(2002)\citenamefont {Bauer},
  \citenamefont {Pirjol},\ and\ \citenamefont {Stewart}}]{Bauer:2001yt}%
  \BibitemOpen
  \bibfield  {author} {\bibinfo {author} {\bibfnamefont {C.~W.}\ \bibnamefont
  {Bauer}}, \bibinfo {author} {\bibfnamefont {D.}~\bibnamefont {Pirjol}}, \
  and\ \bibinfo {author} {\bibfnamefont {I.~W.}\ \bibnamefont {Stewart}},\
  }\href {\doibase 10.1103/PhysRevD.65.054022} {\bibfield  {journal} {\bibinfo
  {journal} {Phys. Rev. D}\ }\textbf {\bibinfo {volume} {65}},\ \bibinfo
  {pages} {054022} (\bibinfo {year} {2002})},\ \Eprint
  {http://arxiv.org/abs/hep-ph/0109045} {hep-ph/0109045} \BibitemShut {NoStop}%
\bibitem [{\citenamefont {Campbell}\ and\ \citenamefont
  {Glover}(1998)}]{Campbell:1997hg}%
  \BibitemOpen
  \bibfield  {author} {\bibinfo {author} {\bibfnamefont {J.~M.}\ \bibnamefont
  {Campbell}}\ and\ \bibinfo {author} {\bibfnamefont {E.~W.~N.}\ \bibnamefont
  {Glover}},\ }\href {\doibase 10.1016/S0550-3213(98)00295-8} {\bibfield
  {journal} {\bibinfo  {journal} {Nucl. Phys.}\ }\textbf {\bibinfo {volume}
  {B527}},\ \bibinfo {pages} {264} (\bibinfo {year} {1998})},\ \Eprint
  {http://arxiv.org/abs/hep-ph/9710255} {arXiv:hep-ph/9710255 [hep-ph]}
  \BibitemShut {NoStop}%
\bibitem [{\citenamefont {Catani}\ and\ \citenamefont
  {Grazzini}(2000)}]{Catani:1999ss}%
  \BibitemOpen
  \bibfield  {author} {\bibinfo {author} {\bibfnamefont {S.}~\bibnamefont
  {Catani}}\ and\ \bibinfo {author} {\bibfnamefont {M.}~\bibnamefont
  {Grazzini}},\ }\href {\doibase 10.1016/S0550-3213(99)00778-6} {\bibfield
  {journal} {\bibinfo  {journal} {Nucl. Phys.}\ }\textbf {\bibinfo {volume}
  {B570}},\ \bibinfo {pages} {287} (\bibinfo {year} {2000})},\ \Eprint
  {http://arxiv.org/abs/hep-ph/9908523} {arXiv:hep-ph/9908523 [hep-ph]}
  \BibitemShut {NoStop}%
\bibitem [{\citenamefont {Baikov}\ \emph {et~al.}(2012)\citenamefont {Baikov},
  \citenamefont {Chetyrkin}, \citenamefont {Kuhn},\ and\ \citenamefont
  {Rittinger}}]{Baikov:2012er}%
  \BibitemOpen
  \bibfield  {author} {\bibinfo {author} {\bibfnamefont {P.~A.}\ \bibnamefont
  {Baikov}}, \bibinfo {author} {\bibfnamefont {K.~G.}\ \bibnamefont
  {Chetyrkin}}, \bibinfo {author} {\bibfnamefont {J.~H.}\ \bibnamefont {Kuhn}},
  \ and\ \bibinfo {author} {\bibfnamefont {J.}~\bibnamefont {Rittinger}},\
  }\href {\doibase 10.1103/PhysRevLett.108.222003} {\bibfield  {journal}
  {\bibinfo  {journal} {Phys. Rev. Lett.}\ }\textbf {\bibinfo {volume} {108}},\
  \bibinfo {pages} {222003} (\bibinfo {year} {2012})},\ \Eprint
  {http://arxiv.org/abs/1201.5804} {arXiv:1201.5804 [hep-ph]} \BibitemShut
  {NoStop}%
\bibitem [{\citenamefont {Herzog}\ \emph {et~al.}(2017)\citenamefont {Herzog},
  \citenamefont {Ruijl}, \citenamefont {Ueda}, \citenamefont {Vermaseren},\
  and\ \citenamefont {Vogt}}]{Herzog:2017dtz}%
  \BibitemOpen
  \bibfield  {author} {\bibinfo {author} {\bibfnamefont {F.}~\bibnamefont
  {Herzog}}, \bibinfo {author} {\bibfnamefont {B.}~\bibnamefont {Ruijl}},
  \bibinfo {author} {\bibfnamefont {T.}~\bibnamefont {Ueda}}, \bibinfo {author}
  {\bibfnamefont {J.~A.~M.}\ \bibnamefont {Vermaseren}}, \ and\ \bibinfo
  {author} {\bibfnamefont {A.}~\bibnamefont {Vogt}},\ }\href {\doibase
  10.1007/JHEP08(2017)113} {\bibfield  {journal} {\bibinfo  {journal} {JHEP}\
  }\textbf {\bibinfo {volume} {08}},\ \bibinfo {pages} {113} (\bibinfo {year}
  {2017})},\ \Eprint {http://arxiv.org/abs/1707.01044} {arXiv:1707.01044
  [hep-ph]} \BibitemShut {NoStop}%
\bibitem [{TZY()}]{TZY:forthcoming}%
  \BibitemOpen
  \href@noop {} {}\bibinfo {howpublished} {T.-Z.~Yang, in
  preparation}\BibitemShut {NoStop}%
\bibitem [{\citenamefont {Del~Duca}\ \emph
  {et~al.}(2016{\natexlab{b}})\citenamefont {Del~Duca}, \citenamefont {Duhr},
  \citenamefont {Kardos}, \citenamefont {Somogyi},\ and\ \citenamefont
  {Tr{\'o}cs{\'a}nyi}}]{DelDuca:2016csb}%
  \BibitemOpen
  \bibfield  {author} {\bibinfo {author} {\bibfnamefont {V.}~\bibnamefont
  {Del~Duca}}, \bibinfo {author} {\bibfnamefont {C.}~\bibnamefont {Duhr}},
  \bibinfo {author} {\bibfnamefont {A.}~\bibnamefont {Kardos}}, \bibinfo
  {author} {\bibfnamefont {G.}~\bibnamefont {Somogyi}}, \ and\ \bibinfo
  {author} {\bibfnamefont {Z.}~\bibnamefont {Tr{\'o}cs{\'a}nyi}},\ }\href
  {\doibase 10.1103/PhysRevLett.117.152004} {\bibfield  {journal} {\bibinfo
  {journal} {Phys. Rev. Lett.}\ }\textbf {\bibinfo {volume} {117}},\ \bibinfo
  {pages} {152004} (\bibinfo {year} {2016}{\natexlab{b}})},\ \Eprint
  {http://arxiv.org/abs/1603.08927} {arXiv:1603.08927 [hep-ph]} \BibitemShut
  {NoStop}%
\bibitem [{\citenamefont {Ferrara}\ \emph {et~al.}(1974)\citenamefont
  {Ferrara}, \citenamefont {Gatto},\ and\ \citenamefont
  {Grillo}}]{Ferrara:1974pt}%
  \BibitemOpen
  \bibfield  {author} {\bibinfo {author} {\bibfnamefont {S.}~\bibnamefont
  {Ferrara}}, \bibinfo {author} {\bibfnamefont {R.}~\bibnamefont {Gatto}}, \
  and\ \bibinfo {author} {\bibfnamefont {A.~F.}\ \bibnamefont {Grillo}},\
  }\href {\doibase 10.1103/PhysRevD.9.3564} {\bibfield  {journal} {\bibinfo
  {journal} {Phys. Rev.}\ }\textbf {\bibinfo {volume} {D9}},\ \bibinfo {pages}
  {3564} (\bibinfo {year} {1974})}\BibitemShut {NoStop}%
\bibitem [{\citenamefont {Mack}(1977)}]{Mack:1975je}%
  \BibitemOpen
  \bibfield  {author} {\bibinfo {author} {\bibfnamefont {G.}~\bibnamefont
  {Mack}},\ }\href {\doibase 10.1007/BF01613145} {\bibfield  {journal}
  {\bibinfo  {journal} {Commun. Math. Phys.}\ }\textbf {\bibinfo {volume}
  {55}},\ \bibinfo {pages} {1} (\bibinfo {year} {1977})}\BibitemShut {NoStop}%
\bibitem [{\citenamefont {Collins}\ and\ \citenamefont
  {Soper}(1981)}]{Collins:1981uk}%
  \BibitemOpen
  \bibfield  {author} {\bibinfo {author} {\bibfnamefont {J.~C.}\ \bibnamefont
  {Collins}}\ and\ \bibinfo {author} {\bibfnamefont {D.~E.}\ \bibnamefont
  {Soper}},\ }\href {\doibase 10.1016/0550-3213(81)90339-4} {\bibfield
  {journal} {\bibinfo  {journal} {Nucl. Phys.}\ }\textbf {\bibinfo {volume}
  {B193}},\ \bibinfo {pages} {381} (\bibinfo {year} {1981})},\ \bibinfo {note}
  {[Erratum: Nucl. Phys.B213,545(1983)]}\BibitemShut {NoStop}%
\bibitem [{\citenamefont {Korchemsky}\ and\ \citenamefont
  {Radyushkin}(1987)}]{Korchemsky:1987wg}%
  \BibitemOpen
  \bibfield  {author} {\bibinfo {author} {\bibfnamefont {G.}~\bibnamefont
  {Korchemsky}}\ and\ \bibinfo {author} {\bibfnamefont {A.}~\bibnamefont
  {Radyushkin}},\ }\href {\doibase 10.1016/0550-3213(87)90277-X} {\bibfield
  {journal} {\bibinfo  {journal} {Nucl.Phys.}\ }\textbf {\bibinfo {volume}
  {B283}},\ \bibinfo {pages} {342} (\bibinfo {year} {1987})}\BibitemShut
  {NoStop}%
\bibitem [{\citenamefont {Banks}\ and\ \citenamefont
  {Zaks}(1982)}]{Banks:1981nn}%
  \BibitemOpen
  \bibfield  {author} {\bibinfo {author} {\bibfnamefont {T.}~\bibnamefont
  {Banks}}\ and\ \bibinfo {author} {\bibfnamefont {A.}~\bibnamefont {Zaks}},\
  }\href {\doibase 10.1016/0550-3213(82)90035-9} {\bibfield  {journal}
  {\bibinfo  {journal} {Nucl. Phys.}\ }\textbf {\bibinfo {volume} {B196}},\
  \bibinfo {pages} {189} (\bibinfo {year} {1982})}\BibitemShut {NoStop}%
\bibitem [{\citenamefont {Gracey}(1994)}]{Gracey:1994nn}%
  \BibitemOpen
  \bibfield  {author} {\bibinfo {author} {\bibfnamefont {J.~A.}\ \bibnamefont
  {Gracey}},\ }\href {\doibase 10.1016/0370-2693(94)90502-9} {\bibfield
  {journal} {\bibinfo  {journal} {Phys. Lett.}\ }\textbf {\bibinfo {volume}
  {B322}},\ \bibinfo {pages} {141} (\bibinfo {year} {1994})},\ \Eprint
  {http://arxiv.org/abs/hep-ph/9401214} {arXiv:hep-ph/9401214 [hep-ph]}
  \BibitemShut {NoStop}%
\bibitem [{\citenamefont {Gardi}(2005)}]{Gardi:2005yi}%
  \BibitemOpen
  \bibfield  {author} {\bibinfo {author} {\bibfnamefont {E.}~\bibnamefont
  {Gardi}},\ }\href {\doibase 10.1088/1126-6708/2005/02/053} {\bibfield
  {journal} {\bibinfo  {journal} {JHEP}\ }\textbf {\bibinfo {volume} {02}},\
  \bibinfo {pages} {053} (\bibinfo {year} {2005})},\ \Eprint
  {http://arxiv.org/abs/hep-ph/0501257} {arXiv:hep-ph/0501257 [hep-ph]}
  \BibitemShut {NoStop}%
\bibitem [{\citenamefont {Belitsky}\ \emph {et~al.}(1999)\citenamefont
  {Belitsky}, \citenamefont {M{\"u}ller},\ and\ \citenamefont
  {Sch{\"a}fer}}]{Belitsky:1998gu}%
  \BibitemOpen
  \bibfield  {author} {\bibinfo {author} {\bibfnamefont {A.~V.}\ \bibnamefont
  {Belitsky}}, \bibinfo {author} {\bibfnamefont {D.}~\bibnamefont
  {M{\"u}ller}}, \ and\ \bibinfo {author} {\bibfnamefont {A.}~\bibnamefont
  {Sch{\"a}fer}},\ }\href {\doibase 10.1016/S0370-2693(99)00146-X} {\bibfield
  {journal} {\bibinfo  {journal} {Phys. Lett.}\ }\textbf {\bibinfo {volume}
  {B450}},\ \bibinfo {pages} {126} (\bibinfo {year} {1999})},\ \Eprint
  {http://arxiv.org/abs/hep-ph/9811484} {arXiv:hep-ph/9811484 [hep-ph]}
  \BibitemShut {NoStop}%
\bibitem [{\citenamefont {Kotikov}\ \emph {et~al.}(2004)\citenamefont
  {Kotikov}, \citenamefont {Lipatov}, \citenamefont {Onishchenko},\ and\
  \citenamefont {Velizhanin}}]{Kotikov:2004er}%
  \BibitemOpen
  \bibfield  {author} {\bibinfo {author} {\bibfnamefont {A.~V.}\ \bibnamefont
  {Kotikov}}, \bibinfo {author} {\bibfnamefont {L.~N.}\ \bibnamefont
  {Lipatov}}, \bibinfo {author} {\bibfnamefont {A.~I.}\ \bibnamefont
  {Onishchenko}}, \ and\ \bibinfo {author} {\bibfnamefont {V.~N.}\ \bibnamefont
  {Velizhanin}},\ }\href {\doibase 10.1016/j.physletb.2004.05.078,
  10.1016/j.physletb.2005.11.002} {\bibfield  {journal} {\bibinfo  {journal}
  {Phys. Lett.}\ }\textbf {\bibinfo {volume} {B595}},\ \bibinfo {pages} {521}
  (\bibinfo {year} {2004})},\ \bibinfo {note} {[Erratum: Phys. Lett. B632, 754
  (2006)]},\ \Eprint {http://arxiv.org/abs/hep-th/0404092}
  {arXiv:hep-th/0404092 [hep-th]} \BibitemShut {NoStop}%
\bibitem [{\citenamefont {Korchemsky}\ and\ \citenamefont
  {Sokatchev}(2015)}]{Korchemsky:2015ssa}%
  \BibitemOpen
  \bibfield  {author} {\bibinfo {author} {\bibfnamefont {G.~P.}\ \bibnamefont
  {Korchemsky}}\ and\ \bibinfo {author} {\bibfnamefont {E.}~\bibnamefont
  {Sokatchev}},\ }\href {\doibase 10.1007/JHEP12(2015)133} {\bibfield
  {journal} {\bibinfo  {journal} {JHEP}\ }\textbf {\bibinfo {volume} {12}},\
  \bibinfo {pages} {133} (\bibinfo {year} {2015})},\ \Eprint
  {http://arxiv.org/abs/1504.07904} {arXiv:1504.07904 [hep-th]} \BibitemShut
  {NoStop}%
\bibitem [{\citenamefont {Belitsky}\ \emph {et~al.}(2016)\citenamefont
  {Belitsky}, \citenamefont {Hohenegger}, \citenamefont {Korchemsky},\ and\
  \citenamefont {Sokatchev}}]{Belitsky:2014zha}%
  \BibitemOpen
  \bibfield  {author} {\bibinfo {author} {\bibfnamefont {A.~V.}\ \bibnamefont
  {Belitsky}}, \bibinfo {author} {\bibfnamefont {S.}~\bibnamefont
  {Hohenegger}}, \bibinfo {author} {\bibfnamefont {G.~P.}\ \bibnamefont
  {Korchemsky}}, \ and\ \bibinfo {author} {\bibfnamefont {E.}~\bibnamefont
  {Sokatchev}},\ }\href {\doibase 10.1016/j.nuclphysb.2016.01.008} {\bibfield
  {journal} {\bibinfo  {journal} {Nucl. Phys.}\ }\textbf {\bibinfo {volume}
  {B904}},\ \bibinfo {pages} {176} (\bibinfo {year} {2016})},\ \Eprint
  {http://arxiv.org/abs/1409.2502} {arXiv:1409.2502 [hep-th]} \BibitemShut
  {NoStop}%
\bibitem [{\citenamefont {Alday}\ \emph {et~al.}(2015)\citenamefont {Alday},
  \citenamefont {Bissi},\ and\ \citenamefont {Lukowski}}]{Alday:2015eya}%
  \BibitemOpen
  \bibfield  {author} {\bibinfo {author} {\bibfnamefont {L.~F.}\ \bibnamefont
  {Alday}}, \bibinfo {author} {\bibfnamefont {A.}~\bibnamefont {Bissi}}, \ and\
  \bibinfo {author} {\bibfnamefont {T.}~\bibnamefont {Lukowski}},\ }\href
  {\doibase 10.1007/JHEP11(2015)101} {\bibfield  {journal} {\bibinfo  {journal}
  {JHEP}\ }\textbf {\bibinfo {volume} {11}},\ \bibinfo {pages} {101} (\bibinfo
  {year} {2015})},\ \Eprint {http://arxiv.org/abs/1502.07707} {arXiv:1502.07707
  [hep-th]} \BibitemShut {NoStop}%
\bibitem [{\citenamefont {Dokshitzer}\ \emph {et~al.}(1996)\citenamefont
  {Dokshitzer}, \citenamefont {Khoze},\ and\ \citenamefont
  {Troian}}]{Dokshitzer:1995ev}%
  \BibitemOpen
  \bibfield  {author} {\bibinfo {author} {\bibfnamefont {Y.~L.}\ \bibnamefont
  {Dokshitzer}}, \bibinfo {author} {\bibfnamefont {V.~A.}\ \bibnamefont
  {Khoze}}, \ and\ \bibinfo {author} {\bibfnamefont {S.~I.}\ \bibnamefont
  {Troian}},\ }\href {\doibase 10.1103/PhysRevD.53.89} {\bibfield  {journal}
  {\bibinfo  {journal} {Phys. Rev.}\ }\textbf {\bibinfo {volume} {D53}},\
  \bibinfo {pages} {89} (\bibinfo {year} {1996})},\ \Eprint
  {http://arxiv.org/abs/hep-ph/9506425} {arXiv:hep-ph/9506425 [hep-ph]}
  \BibitemShut {NoStop}%
\bibitem [{\citenamefont {Kotikov}\ and\ \citenamefont
  {Lipatov}(2007)}]{Kotikov:2006ts}%
  \BibitemOpen
  \bibfield  {author} {\bibinfo {author} {\bibfnamefont {A.~V.}\ \bibnamefont
  {Kotikov}}\ and\ \bibinfo {author} {\bibfnamefont {L.~N.}\ \bibnamefont
  {Lipatov}},\ }\href {\doibase 10.1016/j.nuclphysb.2007.01.020} {\bibfield
  {journal} {\bibinfo  {journal} {Nucl. Phys.}\ }\textbf {\bibinfo {volume}
  {B769}},\ \bibinfo {pages} {217} (\bibinfo {year} {2007})},\ \Eprint
  {http://arxiv.org/abs/hep-th/0611204} {arXiv:hep-th/0611204 [hep-th]}
  \BibitemShut {NoStop}%
\bibitem [{\citenamefont {Kotikov}\ and\ \citenamefont
  {Lipatov}(2003)}]{Kotikov:2002ab}%
  \BibitemOpen
  \bibfield  {author} {\bibinfo {author} {\bibfnamefont {A.~V.}\ \bibnamefont
  {Kotikov}}\ and\ \bibinfo {author} {\bibfnamefont {L.~N.}\ \bibnamefont
  {Lipatov}},\ }\href {\doibase 10.1016/S0550-3213(03)00264-5,
  10.1016/j.nuclphysb.2004.02.032} {\bibfield  {journal} {\bibinfo  {journal}
  {Nucl. Phys.}\ }\textbf {\bibinfo {volume} {B661}},\ \bibinfo {pages} {19}
  (\bibinfo {year} {2003})},\ \bibinfo {note} {[Erratum: Nucl.
  Phys.B685,405(2004)]},\ \Eprint {http://arxiv.org/abs/hep-ph/0208220}
  {arXiv:hep-ph/0208220 [hep-ph]} \BibitemShut {NoStop}%
\bibitem [{\citenamefont {Kotikov}\ \emph {et~al.}(2003)\citenamefont
  {Kotikov}, \citenamefont {Lipatov},\ and\ \citenamefont
  {Velizhanin}}]{Kotikov:2003fb}%
  \BibitemOpen
  \bibfield  {author} {\bibinfo {author} {\bibfnamefont {A.~V.}\ \bibnamefont
  {Kotikov}}, \bibinfo {author} {\bibfnamefont {L.~N.}\ \bibnamefont
  {Lipatov}}, \ and\ \bibinfo {author} {\bibfnamefont {V.~N.}\ \bibnamefont
  {Velizhanin}},\ }\href {\doibase 10.1016/S0370-2693(03)00184-9} {\bibfield
  {journal} {\bibinfo  {journal} {Phys. Lett.}\ }\textbf {\bibinfo {volume}
  {B557}},\ \bibinfo {pages} {114} (\bibinfo {year} {2003})},\ \Eprint
  {http://arxiv.org/abs/hep-ph/0301021} {arXiv:hep-ph/0301021 [hep-ph]}
  \BibitemShut {NoStop}%
\bibitem [{\citenamefont {Bajnok}\ \emph {et~al.}(2009)\citenamefont {Bajnok},
  \citenamefont {Janik},\ and\ \citenamefont {Lukowski}}]{Bajnok:2008qj}%
  \BibitemOpen
  \bibfield  {author} {\bibinfo {author} {\bibfnamefont {Z.}~\bibnamefont
  {Bajnok}}, \bibinfo {author} {\bibfnamefont {R.~A.}\ \bibnamefont {Janik}}, \
  and\ \bibinfo {author} {\bibfnamefont {T.}~\bibnamefont {Lukowski}},\ }\href
  {\doibase 10.1016/j.nuclphysb.2009.02.005} {\bibfield  {journal} {\bibinfo
  {journal} {Nucl. Phys.}\ }\textbf {\bibinfo {volume} {B816}},\ \bibinfo
  {pages} {376} (\bibinfo {year} {2009})},\ \Eprint
  {http://arxiv.org/abs/0811.4448} {arXiv:0811.4448 [hep-th]} \BibitemShut
  {NoStop}%
\bibitem [{\citenamefont {Kotikov}\ \emph {et~al.}(2007)\citenamefont
  {Kotikov}, \citenamefont {Lipatov}, \citenamefont {Rej}, \citenamefont
  {Staudacher},\ and\ \citenamefont {Velizhanin}}]{Kotikov:2007cy}%
  \BibitemOpen
  \bibfield  {author} {\bibinfo {author} {\bibfnamefont {A.~V.}\ \bibnamefont
  {Kotikov}}, \bibinfo {author} {\bibfnamefont {L.~N.}\ \bibnamefont
  {Lipatov}}, \bibinfo {author} {\bibfnamefont {A.}~\bibnamefont {Rej}},
  \bibinfo {author} {\bibfnamefont {M.}~\bibnamefont {Staudacher}}, \ and\
  \bibinfo {author} {\bibfnamefont {V.~N.}\ \bibnamefont {Velizhanin}},\ }\href
  {\doibase 10.1088/1742-5468/2007/10/P10003} {\bibfield  {journal} {\bibinfo
  {journal} {J. Stat. Mech.}\ }\textbf {\bibinfo {volume} {0710}},\ \bibinfo
  {pages} {P10003} (\bibinfo {year} {2007})},\ \Eprint
  {http://arxiv.org/abs/0704.3586} {arXiv:0704.3586 [hep-th]} \BibitemShut
  {NoStop}%
\bibitem [{\citenamefont
  {Velizhanin}(2014{\natexlab{a}})}]{Velizhanin:2014zla}%
  \BibitemOpen
  \bibfield  {author} {\bibinfo {author} {\bibfnamefont {V.~N.}\ \bibnamefont
  {Velizhanin}},\ }\href {\doibase 10.1016/j.nuclphysb.2014.06.021} {\bibfield
  {journal} {\bibinfo  {journal} {Nucl. Phys.}\ }\textbf {\bibinfo {volume}
  {B885}},\ \bibinfo {pages} {772} (\bibinfo {year} {2014}{\natexlab{a}})},\
  \Eprint {http://arxiv.org/abs/1404.7107} {arXiv:1404.7107 [hep-th]}
  \BibitemShut {NoStop}%
\bibitem [{\citenamefont
  {Velizhanin}(2014{\natexlab{b}})}]{Velizhanin:2014fua}%
  \BibitemOpen
  \bibfield  {author} {\bibinfo {author} {\bibfnamefont {V.~N.}\ \bibnamefont
  {Velizhanin}},\ }\href@noop {} {\  (\bibinfo {year} {2014}{\natexlab{b}})},\
  \Eprint {http://arxiv.org/abs/1411.1331} {arXiv:1411.1331 [hep-ph]}
  \BibitemShut {NoStop}%
\bibitem [{\citenamefont {Lukowski}\ \emph {et~al.}(2010)\citenamefont
  {Lukowski}, \citenamefont {Rej},\ and\ \citenamefont
  {Velizhanin}}]{Lukowski:2009ce}%
  \BibitemOpen
  \bibfield  {author} {\bibinfo {author} {\bibfnamefont {T.}~\bibnamefont
  {Lukowski}}, \bibinfo {author} {\bibfnamefont {A.}~\bibnamefont {Rej}}, \
  and\ \bibinfo {author} {\bibfnamefont {V.~N.}\ \bibnamefont {Velizhanin}},\
  }\href {\doibase 10.1016/j.nuclphysb.2010.01.008} {\bibfield  {journal}
  {\bibinfo  {journal} {Nucl. Phys.}\ }\textbf {\bibinfo {volume} {B831}},\
  \bibinfo {pages} {105} (\bibinfo {year} {2010})},\ \Eprint
  {http://arxiv.org/abs/0912.1624} {arXiv:0912.1624 [hep-th]} \BibitemShut
  {NoStop}%
\bibitem [{\citenamefont
  {Velizhanin}(2014{\natexlab{c}})}]{Velizhanin:2013vla}%
  \BibitemOpen
  \bibfield  {author} {\bibinfo {author} {\bibfnamefont {V.~N.}\ \bibnamefont
  {Velizhanin}},\ }\href {\doibase 10.1007/JHEP06(2014)108} {\bibfield
  {journal} {\bibinfo  {journal} {JHEP}\ }\textbf {\bibinfo {volume} {06}},\
  \bibinfo {pages} {108} (\bibinfo {year} {2014}{\natexlab{c}})},\ \Eprint
  {http://arxiv.org/abs/1311.6953} {arXiv:1311.6953 [hep-th]} \BibitemShut
  {NoStop}%
\bibitem [{\citenamefont {Marboe}\ \emph {et~al.}(2015)\citenamefont {Marboe},
  \citenamefont {Velizhanin},\ and\ \citenamefont {Volin}}]{Marboe:2014sya}%
  \BibitemOpen
  \bibfield  {author} {\bibinfo {author} {\bibfnamefont {C.}~\bibnamefont
  {Marboe}}, \bibinfo {author} {\bibfnamefont {V.}~\bibnamefont {Velizhanin}},
  \ and\ \bibinfo {author} {\bibfnamefont {D.}~\bibnamefont {Volin}},\ }\href
  {\doibase 10.1007/JHEP07(2015)084} {\bibfield  {journal} {\bibinfo  {journal}
  {JHEP}\ }\textbf {\bibinfo {volume} {07}},\ \bibinfo {pages} {084} (\bibinfo
  {year} {2015})},\ \Eprint {http://arxiv.org/abs/1412.4762} {arXiv:1412.4762
  [hep-th]} \BibitemShut {NoStop}%
\bibitem [{\citenamefont {Marboe}\ and\ \citenamefont
  {Velizhanin}(2016)}]{Marboe:2016igj}%
  \BibitemOpen
  \bibfield  {author} {\bibinfo {author} {\bibfnamefont {C.}~\bibnamefont
  {Marboe}}\ and\ \bibinfo {author} {\bibfnamefont {V.}~\bibnamefont
  {Velizhanin}},\ }\href {\doibase 10.1007/JHEP11(2016)013} {\bibfield
  {journal} {\bibinfo  {journal} {JHEP}\ }\textbf {\bibinfo {volume} {11}},\
  \bibinfo {pages} {013} (\bibinfo {year} {2016})},\ \Eprint
  {http://arxiv.org/abs/1607.06047} {arXiv:1607.06047 [hep-th]} \BibitemShut
  {NoStop}%
\bibitem [{\citenamefont {Gromov}\ \emph
  {et~al.}(2014{\natexlab{a}})\citenamefont {Gromov}, \citenamefont {Kazakov},
  \citenamefont {Leurent},\ and\ \citenamefont {Volin}}]{Gromov:2013pga}%
  \BibitemOpen
  \bibfield  {author} {\bibinfo {author} {\bibfnamefont {N.}~\bibnamefont
  {Gromov}}, \bibinfo {author} {\bibfnamefont {V.}~\bibnamefont {Kazakov}},
  \bibinfo {author} {\bibfnamefont {S.}~\bibnamefont {Leurent}}, \ and\
  \bibinfo {author} {\bibfnamefont {D.}~\bibnamefont {Volin}},\ }\href
  {\doibase 10.1103/PhysRevLett.112.011602} {\bibfield  {journal} {\bibinfo
  {journal} {Phys. Rev. Lett.}\ }\textbf {\bibinfo {volume} {112}},\ \bibinfo
  {pages} {011602} (\bibinfo {year} {2014}{\natexlab{a}})},\ \Eprint
  {http://arxiv.org/abs/1305.1939} {arXiv:1305.1939 [hep-th]} \BibitemShut
  {NoStop}%
\bibitem [{\citenamefont {Gromov}\ \emph {et~al.}(2015)\citenamefont {Gromov},
  \citenamefont {Kazakov}, \citenamefont {Leurent},\ and\ \citenamefont
  {Volin}}]{Gromov:2014caa}%
  \BibitemOpen
  \bibfield  {author} {\bibinfo {author} {\bibfnamefont {N.}~\bibnamefont
  {Gromov}}, \bibinfo {author} {\bibfnamefont {V.}~\bibnamefont {Kazakov}},
  \bibinfo {author} {\bibfnamefont {S.}~\bibnamefont {Leurent}}, \ and\
  \bibinfo {author} {\bibfnamefont {D.}~\bibnamefont {Volin}},\ }\href
  {\doibase 10.1007/JHEP09(2015)187} {\bibfield  {journal} {\bibinfo  {journal}
  {JHEP}\ }\textbf {\bibinfo {volume} {09}},\ \bibinfo {pages} {187} (\bibinfo
  {year} {2015})},\ \Eprint {http://arxiv.org/abs/1405.4857} {arXiv:1405.4857
  [hep-th]} \BibitemShut {NoStop}%
\bibitem [{\citenamefont {Gromov}\ \emph
  {et~al.}(2014{\natexlab{b}})\citenamefont {Gromov}, \citenamefont
  {Levkovich-Maslyuk}, \citenamefont {Sizov},\ and\ \citenamefont
  {Valatka}}]{Gromov:2014bva}%
  \BibitemOpen
  \bibfield  {author} {\bibinfo {author} {\bibfnamefont {N.}~\bibnamefont
  {Gromov}}, \bibinfo {author} {\bibfnamefont {F.}~\bibnamefont
  {Levkovich-Maslyuk}}, \bibinfo {author} {\bibfnamefont {G.}~\bibnamefont
  {Sizov}}, \ and\ \bibinfo {author} {\bibfnamefont {S.}~\bibnamefont
  {Valatka}},\ }\href {\doibase 10.1007/JHEP07(2014)156} {\bibfield  {journal}
  {\bibinfo  {journal} {JHEP}\ }\textbf {\bibinfo {volume} {07}},\ \bibinfo
  {pages} {156} (\bibinfo {year} {2014}{\natexlab{b}})},\ \Eprint
  {http://arxiv.org/abs/1402.0871} {arXiv:1402.0871 [hep-th]} \BibitemShut
  {NoStop}%
\bibitem [{\citenamefont {Gromov}\ \emph {et~al.}(2016)\citenamefont {Gromov},
  \citenamefont {Levkovich-Maslyuk},\ and\ \citenamefont
  {Sizov}}]{Gromov:2015wca}%
  \BibitemOpen
  \bibfield  {author} {\bibinfo {author} {\bibfnamefont {N.}~\bibnamefont
  {Gromov}}, \bibinfo {author} {\bibfnamefont {F.}~\bibnamefont
  {Levkovich-Maslyuk}}, \ and\ \bibinfo {author} {\bibfnamefont
  {G.}~\bibnamefont {Sizov}},\ }\href {\doibase 10.1007/JHEP06(2016)036}
  {\bibfield  {journal} {\bibinfo  {journal} {JHEP}\ }\textbf {\bibinfo
  {volume} {06}},\ \bibinfo {pages} {036} (\bibinfo {year} {2016})},\ \Eprint
  {http://arxiv.org/abs/1504.06640} {arXiv:1504.06640 [hep-th]} \BibitemShut
  {NoStop}%
\bibitem [{\citenamefont {Kotikov}\ and\ \citenamefont
  {Velizhanin}(2005)}]{Kotikov:2005gr}%
  \BibitemOpen
  \bibfield  {author} {\bibinfo {author} {\bibfnamefont {A.~V.}\ \bibnamefont
  {Kotikov}}\ and\ \bibinfo {author} {\bibfnamefont {V.~N.}\ \bibnamefont
  {Velizhanin}},\ }\href@noop {} {\  (\bibinfo {year} {2005})},\ \Eprint
  {http://arxiv.org/abs/hep-ph/0501274} {arXiv:hep-ph/0501274 [hep-ph]}
  \BibitemShut {NoStop}%
\bibitem [{\citenamefont {Eden}(2012)}]{Eden:2012rr}%
  \BibitemOpen
  \bibfield  {author} {\bibinfo {author} {\bibfnamefont {B.}~\bibnamefont
  {Eden}},\ }\href@noop {} {\  (\bibinfo {year} {2012})},\ \Eprint
  {http://arxiv.org/abs/1207.3112} {arXiv:1207.3112 [hep-th]} \BibitemShut
  {NoStop}%
\bibitem [{\citenamefont {Kotikov}\ and\ \citenamefont
  {Lipatov}(2001)}]{Kotikov:2001sc}%
  \BibitemOpen
  \bibfield  {author} {\bibinfo {author} {\bibfnamefont {A.~V.}\ \bibnamefont
  {Kotikov}}\ and\ \bibinfo {author} {\bibfnamefont {L.~N.}\ \bibnamefont
  {Lipatov}},\ }in\ \href@noop {} {\emph {\bibinfo {booktitle} {{35th Annual
  Winter School on Nuclear and Particle Physics Repino, Russia, February 19-25,
  2001}}}}\ (\bibinfo {year} {2001})\ \Eprint
  {http://arxiv.org/abs/hep-ph/0112346} {arXiv:hep-ph/0112346 [hep-ph]}
  \BibitemShut {NoStop}%
\bibitem [{\citenamefont {Moch}\ \emph {et~al.}(2017)\citenamefont {Moch},
  \citenamefont {Ruijl}, \citenamefont {Ueda}, \citenamefont {Vermaseren},\
  and\ \citenamefont {Vogt}}]{Moch:2017uml}%
  \BibitemOpen
  \bibfield  {author} {\bibinfo {author} {\bibfnamefont {S.}~\bibnamefont
  {Moch}}, \bibinfo {author} {\bibfnamefont {B.}~\bibnamefont {Ruijl}},
  \bibinfo {author} {\bibfnamefont {T.}~\bibnamefont {Ueda}}, \bibinfo {author}
  {\bibfnamefont {J.~A.~M.}\ \bibnamefont {Vermaseren}}, \ and\ \bibinfo
  {author} {\bibfnamefont {A.}~\bibnamefont {Vogt}},\ }\href {\doibase
  10.1007/JHEP10(2017)041} {\bibfield  {journal} {\bibinfo  {journal} {JHEP}\
  }\textbf {\bibinfo {volume} {10}},\ \bibinfo {pages} {041} (\bibinfo {year}
  {2017})},\ \Eprint {http://arxiv.org/abs/1707.08315} {arXiv:1707.08315
  [hep-ph]} \BibitemShut {NoStop}%
\bibitem [{\citenamefont {Vogt}\ \emph {et~al.}(2018)\citenamefont {Vogt},
  \citenamefont {Herzog}, \citenamefont {Moch}, \citenamefont {Ruijl},
  \citenamefont {Ueda},\ and\ \citenamefont {Vermaseren}}]{Vogt:2018miu}%
  \BibitemOpen
  \bibfield  {author} {\bibinfo {author} {\bibfnamefont {A.}~\bibnamefont
  {Vogt}}, \bibinfo {author} {\bibfnamefont {F.}~\bibnamefont {Herzog}},
  \bibinfo {author} {\bibfnamefont {S.}~\bibnamefont {Moch}}, \bibinfo {author}
  {\bibfnamefont {B.}~\bibnamefont {Ruijl}}, \bibinfo {author} {\bibfnamefont
  {T.}~\bibnamefont {Ueda}}, \ and\ \bibinfo {author} {\bibfnamefont
  {J.~A.~M.}\ \bibnamefont {Vermaseren}},\ }\bibfield  {booktitle} {\emph
  {\bibinfo {booktitle} {{Proceedings, 14th DESY Workshop on Elementary
  Particle Physics: Loops and Legs in Quantum Field Theory 2018 (LL2018): St.
  Goar, Germany, April 29-May 04, 2018}}},\ }\href {\doibase
  10.22323/1.303.0050} {\bibfield  {journal} {\bibinfo  {journal} {PoS}\
  }\textbf {\bibinfo {volume} {LL2018}},\ \bibinfo {pages} {050} (\bibinfo
  {year} {2018})},\ \Eprint {http://arxiv.org/abs/1808.08981} {arXiv:1808.08981
  [hep-ph]} \BibitemShut {NoStop}%
\bibitem [{\citenamefont {Beccaria}\ \emph {et~al.}(2007)\citenamefont
  {Beccaria}, \citenamefont {Dokshitzer},\ and\ \citenamefont
  {Marchesini}}]{Beccaria:2007bb}%
  \BibitemOpen
  \bibfield  {author} {\bibinfo {author} {\bibfnamefont {M.}~\bibnamefont
  {Beccaria}}, \bibinfo {author} {\bibfnamefont {{\relax Yu}.~L.}\ \bibnamefont
  {Dokshitzer}}, \ and\ \bibinfo {author} {\bibfnamefont {G.}~\bibnamefont
  {Marchesini}},\ }\href {\doibase 10.1016/j.physletb.2007.07.016} {\bibfield
  {journal} {\bibinfo  {journal} {Phys. Lett.}\ }\textbf {\bibinfo {volume}
  {B652}},\ \bibinfo {pages} {194} (\bibinfo {year} {2007})},\ \Eprint
  {http://arxiv.org/abs/0705.2639} {arXiv:0705.2639 [hep-th]} \BibitemShut
  {NoStop}%
\bibitem [{\citenamefont {Macorini}\ and\ \citenamefont
  {Beccaria}(2010)}]{Macorini:2010px}%
  \BibitemOpen
  \bibfield  {author} {\bibinfo {author} {\bibfnamefont {G.}~\bibnamefont
  {Macorini}}\ and\ \bibinfo {author} {\bibfnamefont {M.}~\bibnamefont
  {Beccaria}},\ }\href@noop {} {\  (\bibinfo {year} {2010})},\ \Eprint
  {http://arxiv.org/abs/1009.5559} {arXiv:1009.5559 [hep-th]} \BibitemShut
  {NoStop}%
\bibitem [{\citenamefont {Belin}\ \emph {et~al.}(2019)\citenamefont {Belin},
  \citenamefont {Hofman},\ and\ \citenamefont {Mathys}}]{Belin:2019mnx}%
  \BibitemOpen
  \bibfield  {author} {\bibinfo {author} {\bibfnamefont {A.}~\bibnamefont
  {Belin}}, \bibinfo {author} {\bibfnamefont {D.~M.}\ \bibnamefont {Hofman}}, \
  and\ \bibinfo {author} {\bibfnamefont {G.}~\bibnamefont {Mathys}},\
  }\href@noop {} {\  (\bibinfo {year} {2019})},\ \Eprint
  {http://arxiv.org/abs/1904.05892} {arXiv:1904.05892 [hep-th]} \BibitemShut
  {NoStop}%
\bibitem [{\citenamefont {Larkoski}\ \emph {et~al.}(2013)\citenamefont
  {Larkoski}, \citenamefont {Salam},\ and\ \citenamefont
  {Thaler}}]{Larkoski:2013eya}%
  \BibitemOpen
  \bibfield  {author} {\bibinfo {author} {\bibfnamefont {A.~J.}\ \bibnamefont
  {Larkoski}}, \bibinfo {author} {\bibfnamefont {G.~P.}\ \bibnamefont {Salam}},
  \ and\ \bibinfo {author} {\bibfnamefont {J.}~\bibnamefont {Thaler}},\ }\href
  {\doibase 10.1007/JHEP06(2013)108} {\bibfield  {journal} {\bibinfo  {journal}
  {JHEP}\ }\textbf {\bibinfo {volume} {1306}},\ \bibinfo {pages} {108}
  (\bibinfo {year} {2013})},\ \Eprint {http://arxiv.org/abs/1305.0007}
  {arXiv:1305.0007 [hep-ph]} \BibitemShut {NoStop}%
\bibitem [{\citenamefont {Larkoski}\ \emph {et~al.}(2014)\citenamefont
  {Larkoski}, \citenamefont {Moult},\ and\ \citenamefont
  {Neill}}]{Larkoski:2014gra}%
  \BibitemOpen
  \bibfield  {author} {\bibinfo {author} {\bibfnamefont {A.~J.}\ \bibnamefont
  {Larkoski}}, \bibinfo {author} {\bibfnamefont {I.}~\bibnamefont {Moult}}, \
  and\ \bibinfo {author} {\bibfnamefont {D.}~\bibnamefont {Neill}},\ }\href
  {\doibase 10.1007/JHEP12(2014)009} {\bibfield  {journal} {\bibinfo  {journal}
  {JHEP}\ }\textbf {\bibinfo {volume} {12}},\ \bibinfo {pages} {009} (\bibinfo
  {year} {2014})},\ \Eprint {http://arxiv.org/abs/1409.6298} {arXiv:1409.6298
  [hep-ph]} \BibitemShut {NoStop}%
\bibitem [{\citenamefont {Larkoski}\ \emph {et~al.}(2016)\citenamefont
  {Larkoski}, \citenamefont {Moult},\ and\ \citenamefont
  {Neill}}]{Larkoski:2015kga}%
  \BibitemOpen
  \bibfield  {author} {\bibinfo {author} {\bibfnamefont {A.~J.}\ \bibnamefont
  {Larkoski}}, \bibinfo {author} {\bibfnamefont {I.}~\bibnamefont {Moult}}, \
  and\ \bibinfo {author} {\bibfnamefont {D.}~\bibnamefont {Neill}},\ }\href
  {\doibase 10.1007/JHEP05(2016)117} {\bibfield  {journal} {\bibinfo  {journal}
  {JHEP}\ }\textbf {\bibinfo {volume} {05}},\ \bibinfo {pages} {117} (\bibinfo
  {year} {2016})},\ \Eprint {http://arxiv.org/abs/1507.03018} {arXiv:1507.03018
  [hep-ph]} \BibitemShut {NoStop}%
\bibitem [{\citenamefont {Moult}\ \emph {et~al.}(2016)\citenamefont {Moult},
  \citenamefont {Necib},\ and\ \citenamefont {Thaler}}]{Moult:2016cvt}%
  \BibitemOpen
  \bibfield  {author} {\bibinfo {author} {\bibfnamefont {I.}~\bibnamefont
  {Moult}}, \bibinfo {author} {\bibfnamefont {L.}~\bibnamefont {Necib}}, \ and\
  \bibinfo {author} {\bibfnamefont {J.}~\bibnamefont {Thaler}},\ }\href
  {\doibase 10.1007/JHEP12(2016)153} {\bibfield  {journal} {\bibinfo  {journal}
  {JHEP}\ }\textbf {\bibinfo {volume} {12}},\ \bibinfo {pages} {153} (\bibinfo
  {year} {2016})},\ \Eprint {http://arxiv.org/abs/1609.07483} {arXiv:1609.07483
  [hep-ph]} \BibitemShut {NoStop}%
\bibitem [{\citenamefont {Komiske}\ \emph {et~al.}(2018)\citenamefont
  {Komiske}, \citenamefont {Metodiev},\ and\ \citenamefont
  {Thaler}}]{Komiske:2017aww}%
  \BibitemOpen
  \bibfield  {author} {\bibinfo {author} {\bibfnamefont {P.~T.}\ \bibnamefont
  {Komiske}}, \bibinfo {author} {\bibfnamefont {E.~M.}\ \bibnamefont
  {Metodiev}}, \ and\ \bibinfo {author} {\bibfnamefont {J.}~\bibnamefont
  {Thaler}},\ }\href {\doibase 10.1007/JHEP04(2018)013} {\bibfield  {journal}
  {\bibinfo  {journal} {JHEP}\ }\textbf {\bibinfo {volume} {04}},\ \bibinfo
  {pages} {013} (\bibinfo {year} {2018})},\ \Eprint
  {http://arxiv.org/abs/1712.07124} {arXiv:1712.07124 [hep-ph]} \BibitemShut
  {NoStop}%
\bibitem [{\citenamefont {Larkoski}\ \emph {et~al.}(2017)\citenamefont
  {Larkoski}, \citenamefont {Moult},\ and\ \citenamefont
  {Nachman}}]{Larkoski:2017jix}%
  \BibitemOpen
  \bibfield  {author} {\bibinfo {author} {\bibfnamefont {A.~J.}\ \bibnamefont
  {Larkoski}}, \bibinfo {author} {\bibfnamefont {I.}~\bibnamefont {Moult}}, \
  and\ \bibinfo {author} {\bibfnamefont {B.}~\bibnamefont {Nachman}},\
  }\href@noop {} {\  (\bibinfo {year} {2017})},\ \Eprint
  {http://arxiv.org/abs/1709.04464} {arXiv:1709.04464 [hep-ph]} \BibitemShut
  {NoStop}%
\end{thebibliography}%
\bibliographystyle{apsrev4-1}

\end{document}